\documentclass[letterpaper,twocolumn,10pt]{article}
\usepackage{hyperref}
\usepackage{usenix2019_v3}

\usepackage{macros}

\usepackage{tikz}
\usepackage{amsmath}

\usepackage{filecontents}

\usepackage{graphicx}
\usepackage{balance}  %
\usepackage{xstring}

\usepackage{listings}
\usepackage{textcomp}

\usepackage{url}
\usepackage{enumitem}

\usepackage{ulem}
\usepackage{multicol}
\usepackage{multirow}
\usepackage{graphicx}
\usepackage{lscape}
\usepackage{algorithm}
\usepackage[noend]{algpseudocode}
\usepackage{subfigure}
\usepackage{amsmath}
\usepackage{amssymb}%
\usepackage{pifont}%

\usepackage{enumitem}
\usepackage{comment}
\usepackage{makecell}
\usepackage[compact]{titlesec}

\usepackage{authblk}
\makeatletter
\renewcommand\AB@affilsepx{, \protect\Affilfont}
\makeatother

\usepackage{tikz}
\usepackage{ragged2e}
\usetikzlibrary{shapes,backgrounds}

\newcommand{\sys}{Cameo\xspace}

\begin{document}

\date{}

\title{Move Fast and Meet Deadlines: Fine-grained Real-time Stream Processing with \sys}

\author[1]{Le Xu\thanks{Contact author: Le Xu <lexu1@illinois.edu>}}
\author[2]{Shivaram Venkataraman}
\author[1]{Indranil Gupta}
\author[3]{Luo Mai}
\author[4]{Rahul Potharaju}
\affil[1]{University of Illinois at Urbana-Champaign}
\affil[2]{UW-Madison}
\affil[3]{University of Edinburgh}
\affil[4]{Microsoft}

\maketitle

\subsection*{Abstract}

Resource provisioning in multi-tenant stream processing systems faces the dual challenges of keeping resource utilization high (without over-provisioning), and ensuring performance isolation. %
In our common production use cases, where streaming workloads have to meet %
latency targets and avoid breaching service-level agreements, existing solutions are incapable of 
handling the wide variability of user needs. 
Our framework called \sys uses fine-grained stream processing (inspired by  actor computation models), and is able to provide  high resource utilization while meeting latency targets. \sys dynamically calculates and propagates priorities of events based on user latency targets and query semantics.  %
Experiments on Microsoft Azure show that compared to state-of-the-art, the \sys framework: i)  reduces query latency by %
2.7$\times$%
in single tenant settings, ii) reduces query latency by
4.6$\times$%
in multi-tenant scenarios, and iii) weathers transient spikes of workload.

\section{Introduction}
\label{sec:intro}

Stream processing applications in large companies handle tens of millions of events per second~\cite{abraham2013scuba, mai2018chi, 10.5555/3323234.3323252}. In an attempt to scale and keep total cost of ownership (TCO) low, today's systems: a) parallelize operators %
across machines, and b) use multi-tenancy, wherein %
operators are collocated on shared resources. 
Yet, resource provisioning in production environments remains challenging due to two major reasons:

\noindent\textbf{(i) High workload variability.} 
In a production cluster at a large online services company, we observed orders of magnitude variation in event ingestion and processing rates, {\it across time}, {\it across data sources},  {\it across operators}, and  {\it across applications}. 
This indicates that resource allocation needs to be dynamically tailored towards each operator in each query, in a nimble and adept manner at run time. 

\noindent\textbf{(ii) Latency targets vary across applications.} User expectations come in myriad shapes.  
Some applications require quick responses to events of interest, i.e., short end-to-end latency. Others wish to maximize throughput under limited resources, and yet others desire high resource utilization.
Violating such user expectations is expensive, resulting in breaches of service-level agreements (SLAs), monetary losses, and  customer dissatisfaction.

To address these challenges, we explore a new {\it fine-grained} philosophy for designing a multi-tenant stream processing system. Our key idea is to provision resources to each {operator} based solely on its {\it immediate} need. 
{Concretely we focus on deadline-driven needs. }%
Our fine-grained approach is inspired by the recent emergence %
of event-driven data processing architectures including %
 actor frameworks like Orleans~\cite{bykov2011orleans, orleans_dotnet} and Akka~\cite{akka}, and serverless cloud platforms~\cite{azure_function, amazon-serverless-streaming, google_cloud_functions, jonas2019cloud}.

Our motivation for exploring a fine-grained approach is to %
{enable resource sharing directly among operators}.  This is more efficient than the traditional {\it slot-based} approach, 
{ wherein operators are assigned dedicated resources. %
}
In the slot-based approach, operators are mapped onto processes or threads---examples include task slots in Flink~\cite{flink}, instances in Heron~\cite{heron}, and executors in Spark Streaming~\cite{sparkstreaming}. 
Developers then need to either assign applications to a dedicated subset of machines\cite{storm_multitenant}, or place execution slots in resource containers and acquire physical resources (CPUs and memory) through resource managers~\cite{yarn, apache_mesos, kubernetes}.

\begin{figure}
\centering
\includegraphics[width=.48\textwidth]{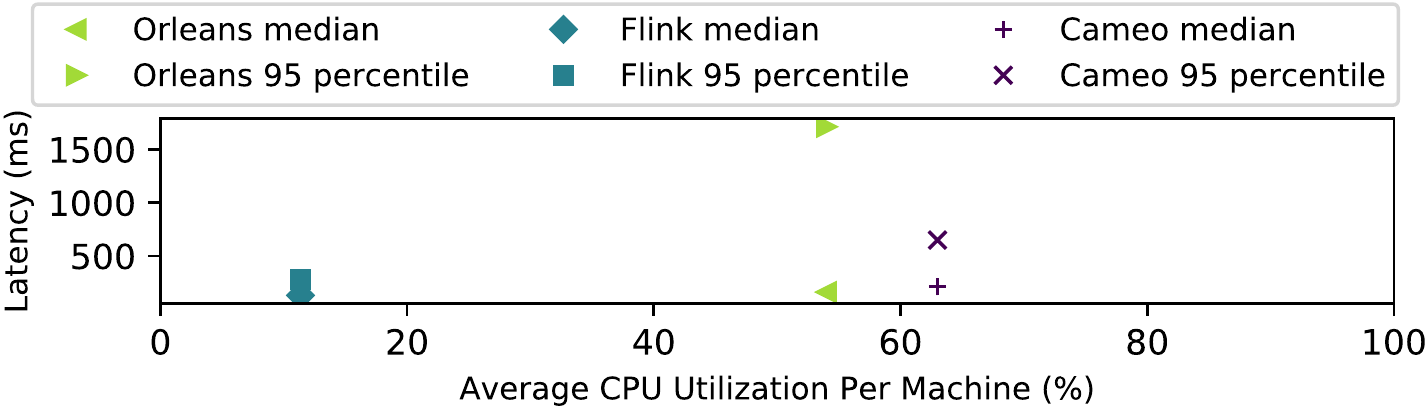}
\vspace{-0.2in}
\caption{\textit{Slot-based system (Flink), Simple Actor system (Orleans), and our framework  \sys.}}
\label{fig:latency-util-tradeoff}
\vspace{-0.2in}
\end{figure}

While slot-based systems provide isolation, they are hard to dynamically reconfigure in the face of workload variability. As a result it has become common for developers to ``game'' their  resource requests, asking for over-provisioned resources, far above what the job needs~\cite{dhalion2017}. Aggressive users starve other jobs which might need immediate resources, and the upshot is unfair allocations and low utilization. 

At the same time, today's fine-grained scheduling systems like Orleans, as shown in
Figure~\ref{fig:latency-util-tradeoff}, cause high tail latencies.
The figure also shows that a slot-based system (Flink on YARN), which maps each executor to a CPU, leads to low  resource utilization. %
The plot shows that our approach, \sys, can provide both high utilization and low tail latency.

To realize our approach, we %
develop a new priority-based framework for fine-grained distributed stream processing. 
This requires us to tackle several {\it architectural} design  challenges including:
1) translating a job's performance target (deadlines) to priorities of individual messages, 2)
developing interfaces to use real-time scheduling policies such as earliest deadline first (EDF)~\cite{liu1973scheduling}, least laxity first (LLF)~\cite{mok1983fundamental} etc., 
and 3) low-overhead scheduling of operators for prioritized  messages.
We present {\it \sys}, a new 
scheduling framework designed for data streaming applications. \sys:
\squishlist
\item {\it Dynamically} derives priorities of operators, using both: a) {\it static input}, e.g., job deadline; and b) {\it dynamic stimulus}, e.g., tracking stream progress, profiled message execution times. 
\item Contributes new mechanisms: i) {\it scheduling contexts}, which propagate scheduling states along dataflow paths,  ii) a {\it context handling} interface, which enables pluggable scheduling strategies (e.g., laxity, deadline, etc.), %
and iii) tackles required scheduling issues including  per-event synchronization, and semantic-awareness to events.
\item Provides low-overhead scheduling
by: i) using a stateless scheduler,  and 
ii) allowing scheduling operations to be  %
driven purely by message arrivals and flow.

\squishend

We build Cameo on Flare~\cite{mai2018chi}, which is a distributed data flow runtime built atop Orleans~\cite{bykov2011orleans, orleans_dotnet}. %
Our experiments are run on Microsoft Azure, using production workloads.  %
\sys,  
using a laxity-based scheduler, 
reduces latency by up to 2.7$\times$ in single-query scenarios and up to 4.6$\times$ in multi-query scenarios. \sys schedules are resilient to transient workload spikes and ingestion rate skews across sources. \sys's  scheduling decisions incur  
less than 6.4\% overhead.

\section{Background and Motivation}
\label{sec:motivation}

\subsection{Workload Characteristics}
\label{sub:workload_features}

\begin{figure}[!bt]
\centering
\begin{tabular}{cc}

\hspace{-.2in}
\subfigure[\textit{Data Volume Distribution}]{
  \includegraphics[width=0.15\textwidth]{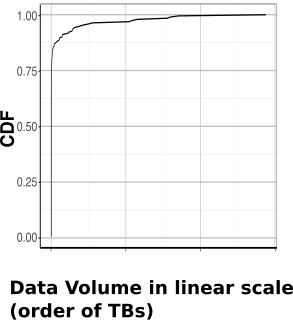}
\label{fig:volume_distribution}
}
\hspace{-.15in}
& 
\subfigure[\textit{Job Scheduling \& Completion Latencies}]{
  \includegraphics[width=0.3\textwidth]{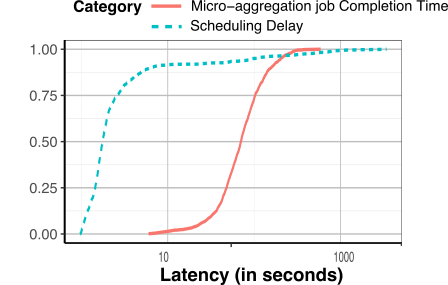}
\label{fig:job_latencies}
}
 \\
 
\multicolumn{2}{c}{
\hspace{-.2in}
\subfigure[\textit{Ingestion Heatmap}]{
    \includegraphics[width=0.45\textwidth]{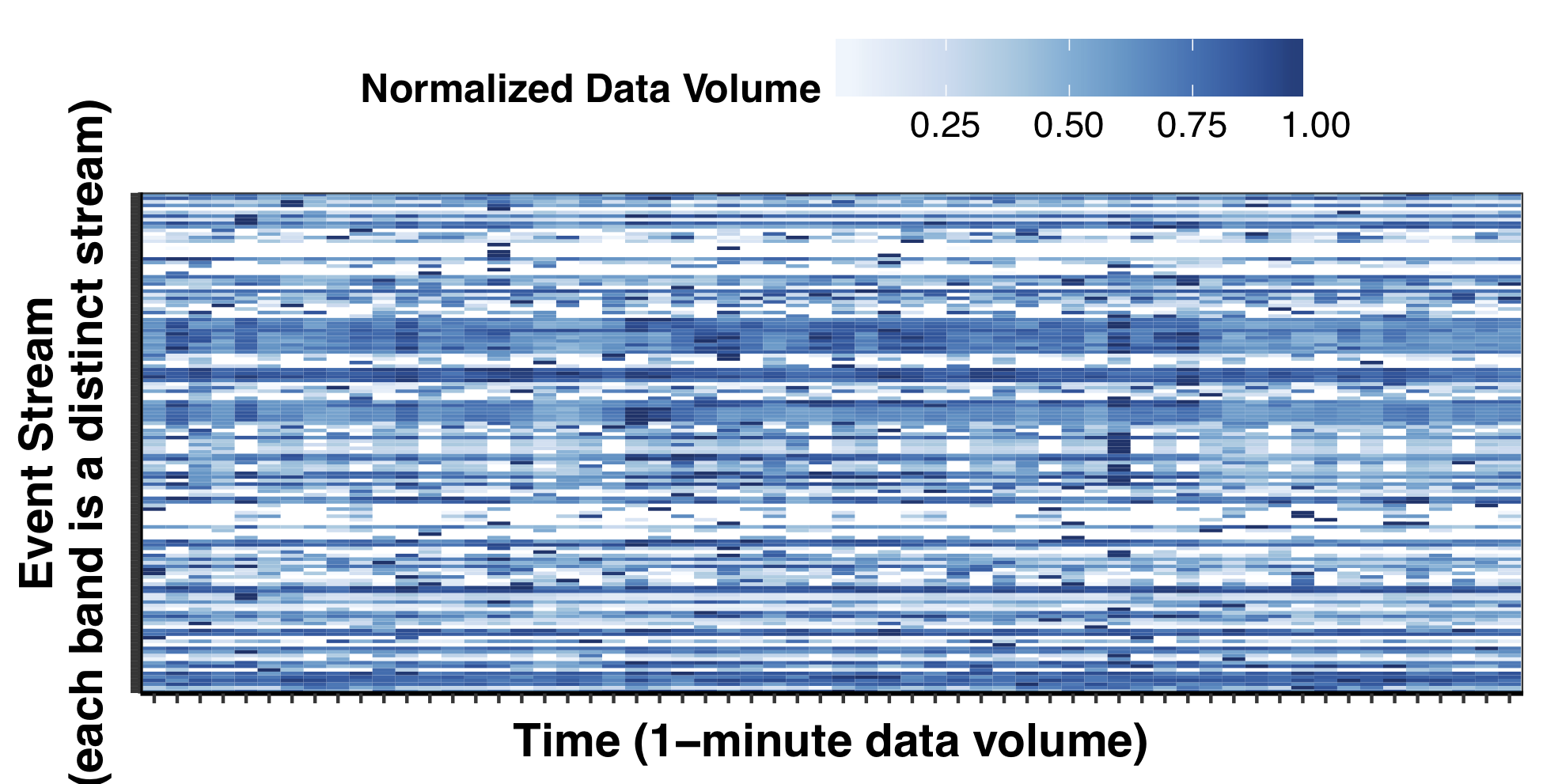}
\label{fig:volume_heatmap}
} 
}
\end{tabular}
\caption{\textit{Workload characteristics collected from a production stream analytics system.} 
}
\vspace{-.2in}
\label{fig:related-work-map}
\end{figure}

We study a production cluster that ingests more than 10 PB per day over several 100K machines. The shared cluster has several internal teams running streaming applications which  perform  debugging, monitoring, impact analysis, etc. We first make key observations about this workload. 

\noindent \textbf{Long-tail streams drive resource over-provisioning.}
Each data stream is handled by a  standing  
\textit{streaming} query, deployed as a dataflow job. 
As shown in Figure~\ref{fig:volume_distribution}, we first observe that 10\% of the streams process a majority of the data. Additionally, 
we observe that a long tail of streams, each processing small amount data, are responsible for over-provisioning---their users rarely have any means of accurately gauging how many nodes are required, and end up over-provisioning for their job.

\noindent \textbf{Temporal variation makes resource prediction difficult.} %
Figure~\ref{fig:volume_heatmap} is a heat map showing incoming data volume for 20 different stream sources. The graph shows a high degree of  variability across both sources and time. A single stream can have spikes lasting one to a few seconds, as well as periods of idleness. Further, this pattern is continuously changing. This points to the need for an agile and fine-grained way to respond to temporal variations, as they are occurring.

\noindent \textbf{Users already try to do fine-grained scheduling.} 
We have observed that instead of continuously running streaming applications, our users prefer to provision a cluster using external resource managers (e.g., YARN~\cite{apache_hadoop}, Mesos~\cite{apache_mesos}), and then run periodic micro-batch jobs. Their implicit aim is to improve resource utilization and  throughput (albeit with unpredictable latencies). However, Figure~\ref{fig:job_latencies} shows that %
this ad-hoc approach causes overheads as high as 80\%. This points to the need for a common way to allow all users to perform fine-grained scheduling, without a hit on performance. %

\noindent \textbf{Latency requirements vary across jobs.}
Finally, we also see a wide range of latency requirements across jobs. 
Figure~\ref{fig:job_latencies} 
shows that the job completion time for the micro-aggregation jobs ranges from less than 10 seconds up to 1000 seconds.  %
This suggests that the range of SLAs required by queries will  vary across a wide range.
This also presents an opportunity for priority-based scheduling: applications have longer latency constraints tend to have greater flexibility in terms of \textit{when} its input can be processed (and vice versa).

\subsection{Prior Approaches}

\noindent\textbf{Dynamic resource provisioning for stream processing.} Dynamic resource provisioning
for streaming data has been addressed %
primarily from the perspective of dataflow reconfiguration. 
 These works fall into three categories as shown in Figure~\ref{fig:related_work_venn}:\\
i) {\it Diagnosis And Policies}: Mechanisms for when and how resource re-allocation is performed; \\
ii) {\it  Elasticity Mechanisms}: Mechanisms for efficient query reconfiguration; and \\
iii) {\it Resource Sharing}: Mechanisms for  dynamic performance isolation among streaming queries.\\ 
These techniques make changes to the dataflows in reaction to a performance metric (e.g., latency) deteriorating.

\sys's approach 
does not involve changes to the dataflow. It is based on the insight that the streaming engine %
 can delay processing of 
those query operators %
which will not violate performance targets right away. %
This allows us to quickly prioritize and provision resources proactively for 
those other operators which could immediately need resources.
At the same time, existing reactive techniques from Figure~\ref{fig:related_work_venn} are orthogonal to our approach and can be used alongside our proactive techniques.

\def\firstcircle{(150:1.2) ellipse (2.4 and 0.8)}
\def\secondcircle{(30:1.2) ellipse (2.4 and 0.8)}
\def\thirdcircle{(-90:0.3) ellipse (2.4 and 0.8)}
\definecolor{mycolor1}{RGB}{173,214,85}
\definecolor{mycolor2}{RGB}{66,126,140}
\definecolor{mycolor3}{RGB}{61,10,81}
\begin{figure}
\begin{tikzpicture}

\draw[mycolor2, line width=0.7mm] \firstcircle;
\draw[mycolor1, line width=0.7mm ] \secondcircle;
\draw[mycolor3, line width=0.7mm] \thirdcircle;

\tikzstyle{bag} = [align=left]
\node at (0,-1.4) [bag] {
\textbf{Elasticity Mechanisms}
};
\node at (-2,0.8)
[font=\small,
  text width=1.8cm]
  {~\cite{li2015supporting,heinze2015online,hoffmann2018snailtrail, kalavri2018three,xu2016stela, hoffmann2019megaphone}};
\node at (-.9, 0)
[font=\small,
  text width=2cm]
  {~\cite{schneider2009elastic,gedik2014elastic}};

\tikzstyle{bag} = [align=right]
\node at (2,1.6) [bag]{
\textbf{Resource Sharing}
};

\node at (2.4,0.8) 
[font=\small,
  text width=2cm]
{~\cite{dhalion2017, garefalakis2018medea}};

\node at (0.5,1) 
[font=\small,
  text width=2cm]
{~\cite{kalim2018henge}};

\tikzstyle{bag} = [align=left]
\node at (-2,1.6) [bag] {
\textbf{Diagnosis And Policies}
};

\node at (0,-0.7)
[font=\small,
  text width=2.2cm]
  {~\cite{castro2013integrating,lohrmann2015elastic, fu2015drs, kalyvianaki2012overload,kalyvianaki2016themis, gulisano2012streamcloud, venkataraman2017drizzle,heinze2014latency, mai2018chi}};
  
\node at (1.8,0)
[font=\small,
  text width=2cm]
  {~\cite{ garefalakis2019neptune}};

\end{tikzpicture}
\caption{
\textit{
Existing Dataflow Reconfiguration Solutions.
} 
}
\label{fig:related_work_venn}
\vspace{-.2in}
\end{figure}
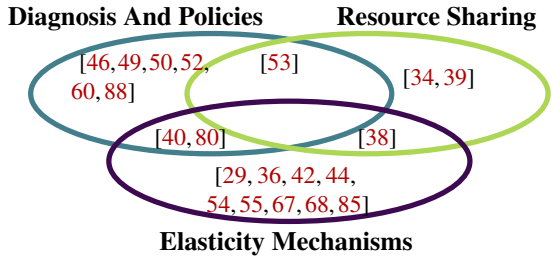

\noindent\textbf{The promise of event-driven systems.} To achieve fine-grained scheduling, a promising direction
is to leverage emerging event-driven systems such as actor frameworks~\cite{newell2016optimizing, haller2009scala}
and serverless platforms~\cite{bernstein2019serverless}. 
Unlike slot-based stream processing systems like Flink~\cite{flink} and
Storm~\cite{storm}, operators here are not mapped to specific CPUs. Instead event-driven systems maintain centralized {queues} to host incoming messages and dynamically
dispatch messages to available CPUs.
This provides an
opportunity to develop systems that can manage a unified queue of messages
across query boundaries, and  %
combat the over-provisioning of slot-based approaches. %
Recent proposals for this execution model also include~\cite{bernstein2019serverless, amazon-serverless-streaming, statefun, kumar2020amber}. 

\sys builds on the rich legacy of work from two communities: classical real-time systems~\cite{li2013deadline, ou2005tick} and first-generation stream management systems (DSMS) in the database community~\cite{abadi2003aurora, abadi2005design, chandrasekaran2003telegraphcq,motwani2003query}. The former category has produced rich scheduling algorithms, but unlike \sys, none build a full working system that is flexible in policies, or support streaming operator semantics.
In the latter category the closest to our work are event-driven approaches~\cite{carney2003operator, abadi2005design, babcock2003chain}. But these do not interpret stream progress to derive priorities or support trigger analysis
for distributed, user-defined operators. Further, they adopt a centralized, stateful scheduler design, where the scheduler \textit{always} maintains state for all queries, making them challenging to scale. %

Achieving \sys's goal of dynamic resource provisioning  is challenging. 
Firstly, messages sent by user-defined operators are a black-box to 
{event} schedulers. Inferring their impact on query {performance} requires new techniques %
to analyze and re-prioritize said messages. %
Secondly, event-driven %
schedulers must scale with message volume and not bottleneck.%

\section{Design Overview}
\label{sec:overview}

\noindent\textbf{Assumptions, System Model:} We design \sys to support streaming queries on clusters shared by %
cooperative users, e.g., within an %
organization. We also assume that the user  specifies a latency target at query submission time, e.g., %
derived from product and service requirements.%

The architecture of \sys consists of two major components: (i) a scheduling strategy which determines message priority by interpreting the semantics of query and data streams given a latency target.
(Section~\ref{sec:scheduling}), and (ii) a scheduling framework that 1. enables message priority to be generated using a pluggable strategy, and 2. schedules operators dynamically based on their current pending messsages' priorities
(Section~\ref{sec:architecture}). %

\sys prioritizes operator processing by computing the {\it start deadlines} of arriving messages, i.e., latest time for a message to start execution at an operator without violating the downstream dataflow's latency target for that message. \sys continuously reorders operator-message pairs to prioritize messages with earlier deadlines.

Calculating priorities requires the scheduler to continuously book-keep both: (i) per-job static information, e.g., latency constraint/requirement\footnote{We use latency constraint and latency requirement interchangeably.} and dataflow topology, and (ii) dynamic information such as the timestamps of tuples being processed (e.g.,  stream progress \cite{li2005semantics, Akidau2015TheDM}), and estimated execution cost per operator. 
To scale such a fine-grained scheduling approach to a large number of jobs, \sys utilizes \textit{scheduling contexts}--- data structures attached to messages that capture and transport information required to generate priorities.

The scheduling framework of \sys has two levels.  The upper level consists of \textit{context converters}, embedded into each operator. A context converter modifies and propagates scheduling contexts attached to a message.
The lower level is a \textit{stateless scheduler} that determines target operator's priority by interpreting scheduling context attached to the message. We also design a programmable API for a pluggable scheduling strategy that can be used to handle scheduling contexts. 
In summary, these design decisions make our scheduler scale to a large number of jobs with low overhead.%

\begin{figure*}
\centering

\includegraphics[width=1\textwidth]{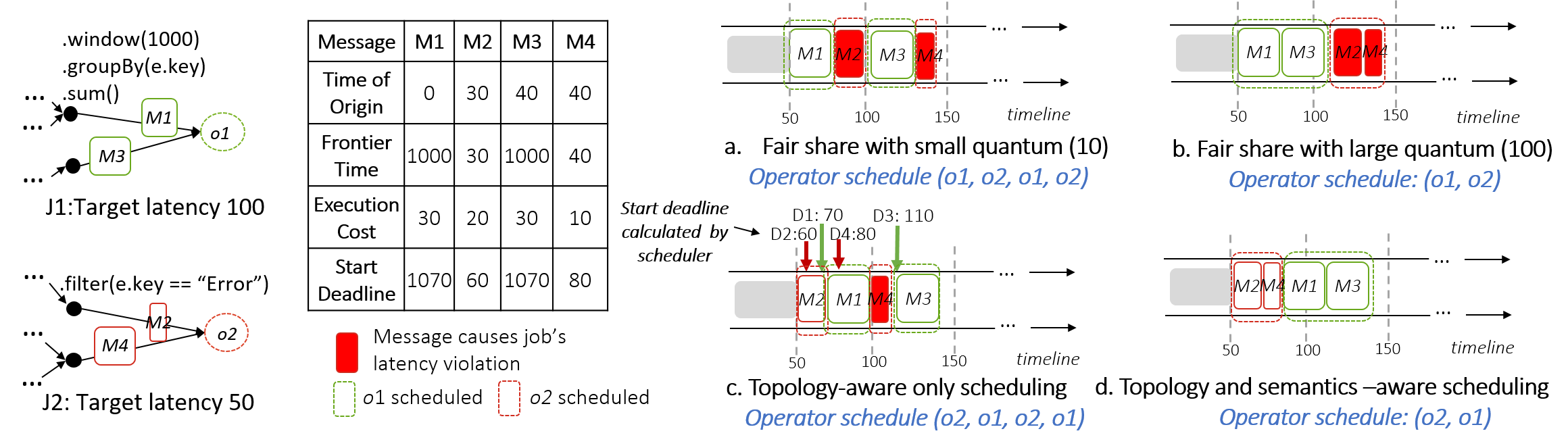}
\vspace{-.2in}
\caption{\textit{Scheduling Example: J1 is batch analytics, J2 is latency-sensitive. Fair-share scheduler creates schedules  ``a'' and ``b''. Topology-aware scheduler reduces  violations  (``c''). Semantics-aware scheduler further reduces violations (``d''). We further explain these examples in Section~\ref{subsec:scenarios}
}}
\label{fig:priority-example}
\vspace{-.1in}
\end{figure*}

\mypar{Example}  We  present an example highlighting our approach.
Consider a workload,  shown in Figure~\ref{fig:priority-example}, consisting of two streaming dataflows $J1$ and $J2$ where $J1$ performs a batch analytics query and $J2$ performs a latency sensitive anomaly detection pipeline. Each has an output operator processing messages from upstream operators.
The default approach used by actor systems like Orleans 
is to: i) order messages based on arrival,  and ii) give each operator a fixed time duration (called ``quantum'') to process its messages. %
Using this approach we derive the schedule ``a'' with a small quantum, and a schedule ``b'' with a large quantum ---
both result in two 
latency violations for $J2$. 
In comparison,  \sys
discovers the opportunity to postpone less latency-sensitive messages (and thus their target operators). This helps $J2$ meet its deadline by leveraging topology and query semantics. This is depicted in schedules ``c'' and ``d''. 
This example shows that \textit{when} and \textit{how long} an operator is scheduled to run should be {dynamically} determined by the priority of the  {{ next} pending message}. We expand on these aspects in the forthcoming sections.

\section{Scheduling Policies in \sys}

\label{sec:scheduling}

One of our primary goals in \sys is to enable fine-grained scheduling policies for dataflows. These policies can prioritize messages based on information, like the deadline remaining or processing time for each message, etc. To enable such policies, we require techniques that can calculate the priority of a message for a given policy.

We model our setting as a non-preemptive, non-uniform task time, multi-processor, real-time %
scheduling problem. Such problems are known to be NP-Complete offline and cannot be solved optially online without complete knowledge of future tasks~\cite{coffman1978application,
stankovic1995implications}. 
Thus, we consider how a number of commonly used policies in this domain, including  Least-Laxity-First (LLF)~\cite{mok1983fundamental}, Earliest-Deadline-First (EDF)~\cite{liu1973scheduling} and Shortest-Job-First (SJF)~\cite{tanenbaum2015modern}, and describe how such policies can be used for event-driven stream processing.%
We use the LLF policy as the default policy in our description below.%

The above policies try to prioritize messages to avoid violating latency constraints.
Deriving the priority of a message requires analyzing the impact of each operator in the dataflow on query performance. We next discuss how being deadline-aware can help \sys derive appropriate priorities. We also discuss how being aware of query semantics can further improve prioritization.

\begin{table}[htb!]
\centering
\resizebox{.42\textwidth}{!}{
\begin{minipage}{.45\textwidth}
\begin{center}
\begin{tabular}{ |c|c| } 
 \hline
 Symbol & Definition  \\ 
 \hline
 $ID_{M}$ &  ID of Message M.\\ 
 $ddl_{M}$ & Message start deadline. \\ 
 $o_{M}$ & target operator of $M$.  \\ 
 $C_{o_{M}}$ & \makecell{Estimated execution cost of $M$ \\on its target operator.} \\
 $t_{M}$, and $p_M$ & \makecell{Physical (and logical) time associated  \\with the last event required to produce $M$.} \\
 $L$ & \makecell{Dataflow latency constraint of the\\ dataflow that $M$ belongs to.} \\
 $p_{M_F}$, and $t_{M_F}$ & \makecell{Frontier progress, and frontier time.}\\
 \hline
\end{tabular} 
\caption{\textit{Notations used in paper for message  $M$.}} 
\label{tbl:symbols}
\end{center}
\end{minipage} }
\end{table}

\subsection{Definitions and Underpinnings}
\label{subsec:definition}

\mypar{Event} Input data arrives as \textit{events}, associated with a \textit{logical time}~\cite{chandramouli2014trill} that indicates the \textit{stream progress} of these events in the input stream.

\mypar{Dataflow job and operators} 
A dataflow job consists of a DAG of  {\it stages}. Each stage operates a user-defined function. A stage can be parallelized and executed by a set of dataflow {\it operators}.

We say an operator $o_{k}$ is {\it invoked} when it processes its input message, and $o_{k}$ is {\it triggered} when it is invoked and leads to an output message, which is either passed downstream to further operators or the final job output.

\sys considers two types of operators: i) regular operators that are triggered immediately on invocation; 
and ii) windowed operators~\cite{li2005semantics} that partitions data stream into sections by logical times and triggers only when all data from the section are observed.

\mypar{Message timestamps} 
We denote a message $M$ as a tuple $(o_{M}, (p_{M}, t_{M}))$, where: a) $o_{M}$ is the operator executing the message; b) $p_{M}$ and $t_{M}$ record the \textit{logical} and \textit{physical} time of the input stream
that is associated with $M$, respectively. 
Intuitively, $M$ is influenced by 
input stream with logical time $\leq$ 
$p_{M}$. Physical time $t_{M}$ marks the system time when $p_{M}$ is observed at a source operator. 

We denote $C_{o_{M}}$ as the estimated time  to process message $M$ on  target operator $O$, and $L$ as the latency constraint for the dataflow that $M$ belongs to.

\mypar{Latency} Consider a message $M$ generated as the output of a dataflow (at its sink operator). Consider the set of all events $E$ that influenced the generation of $M$. We define latency as the difference between the last arrival time of any event in $E$ and the time when $M$ is generated.

\subsection{Calculating Message Deadline}
\label{subsec:scenarios}
We next consider the LLF scheduling policy where we wish to prioritize messages which have the least laxity (i.e.,  flexibility). Intuitively, this allows us to prioritize messages that are closer to violating their latency constraint.
To do this, we  discuss how to determine the {\it latest time} that a message $M$ can start executing at operator $O$ without violating the job's latency constraint. We call this as the {\it start deadline} or in short the {\it deadline} of the  message $M$, denoted as $ddl_{M}$. For the LLF scheduler, $ddl_{M}$ is the message priority (lower value implies higher priority).  %
 
We describe how to derive the priority (deadline) using topology-awareness and then query (semantic)-awareness.

\subsubsection{Topology Awareness}
\label{subsubsec:latency_constraint_awareness}

\mypar{Single-operator dataflow, Regular operator} Consider a dataflow with only one regular operator $o_{M}$. 
The latency constraint is $L$. If an event occurs at time $t_{M}$, then $M$ should complete processing before $t_{M} + L$. The start deadline, given execution estimate $C_{o_{M}}$, is:

\vspace{-0.4cm}
\begin{center}
\begin{equation} 
\label{eq:ddl-1}
ddl_{M} = t_{M} + L - C_{o_{M}}
\end{equation}
\end{center}
\vspace{-0.2cm}

\mypar{Multiple-operator dataflow, Regular operator} For an operator $o$ inside a dataflow DAG that is invoked by message $M$, the start deadline of $M$ needs to account for execution time of downstream operators. We estimate the maximum of execution times of critical path~\cite{hoffmann2018snailtrail} from $o$ to any output operator as $C_{path}$. 
The start deadline of $M$ is then: 
\vspace{-0.5cm}
\begin{center}
\begin{equation} 
\label{eq:ddl-2}
ddl_{M} = t_{M} + L - %
C_{O_M}-C_{path}
 \end{equation}
\end{center}

Schedule ``c'' of Figure~\ref{fig:priority-example} showed an example 
of topology-aware scheduling and how topology awareness helps reduce violations. For example, $ddl_{M2}=30 + 50 - 20 = 60$ means that $M2$ is promoted due to its urgency. 
We later show %
that even when query semantics are not available (e.g., UDFs), \sys improves scheduling with topology information alone.
Note that upstream operators are not involved in this calculation. 
 $C_{O_M}$ and $C_{path}$  can be calculated by profiling.

\subsubsection{Query Awareness}
\label{subsubsec:job_semantic_awareness}

\sys can also leverage dataflow semantics, i.e., knowledge of user-specified commands inside the operators.
This enables the scheduler to identify messages which can tolerate further delay without violating  latency constraints. This is common for windowed operations, e.g., a \texttt{WindowAggregation} operator 
can tolerate delayed execution if a message's logical time is at the start of the window as the operator will only produce output at the end of a window. Window operators are very common in our production use cases.

\mypar{Multiple-operator dataflow, Windowed operator} 
Consider $M$ that targets a windowed operator $o_{M}$, \sys is able to determine (based on dataflow semantics) to what extent $M$ can be delayed without affecting latency. 
This requires \sys to identify the minimum logical time ($p_{M_{F}}$) required to trigger the target window operator. We call $p_{M_{F}}$ \textit{frontier progress}.
Frontier progress denotes the stream progress that needs to be observed at the window operator before a window is complete. Thus a windowed operator will not produce output until frontier progresses are observed at all source operators.  
We record the system time when all frontier progresses become available at all sources as \textit{frontier time}, denoted as $t_{M_{F}}$.

Processing of a message $M$ can be safely delayed until all the messages that belong in the window have arrived. %
In other words when computing the start deadline of $M$, we can extend the deadline by $(t_{M_{F}} - t_{M})$. We thus rewrite Equation~\ref{eq:ddl-2} as:

\vspace{-0.5cm}
\begin{center}
\begin{equation} 
\label{eq:ddl-3}
ddl_{M} = \mathbf{t_{M_{F}}} + L - %
C_{O_M}-C_{path}
\end{equation}
\end{center}

An example of this schedule was shown in schedule ``d'' of Figure~\ref{fig:priority-example}. With query-awareness, scheduler derives $t_{M_{F}}$ and postpones $M1$ and $M3$ in favor of $M2$ and $M4$. Therefore operator $o2$ is prioritized over $o1$ to process $M2$ then $M4$.

The above examples show the derivation of priority for a LLF scheduler. \sys{} also supports scheduling policies including commonly used policies like EDF, SJF etc. In fact, the priority for EDF can be derived by a simple modification of the LLF equations. Our EDF policy considers the deadline of a message prior to an operator executing and thus we can compute priority for EDF by omitting $C_{O_M}$ term in Equation~\ref{eq:ddl-3}. %
For SJF we can derive the priority by setting $ddl_M=C_{O_M}$---while SJF is not deadline-aware we compare its performance to other policies in our evaluation. %

\subsection{Mapping Stream Progress}
\label{subsec:propagation}

For Equation~\ref{eq:ddl-3} frontier time  $t_{M_{F}}$ may not be available until the target operator is triggered. %
However, for many fixed-sized window operations (e.g., \texttt{SlidingWindow}, \texttt{TumblingWindow}, etc.), we can {\it estimate} $t_{M_{F}}$ based on the  message's logical time $p_{M}$. \sys performs two steps: first we apply a \textsc{Transform} function to calculate $p_{M_{F}}$, the logical time of the message that triggers $o_{M}$. Then, \sys infers the frontier time $t_{M_{F}}$ using a \textsc{ProgressMap} function. Thus $t_{M_{F}}$ = \textsc{ProgressMap}(\textsc{Transform}($p_{M}$)). We elaborate below.

\noindent\textbf{Step 1 (Transform):} 
For a windowed operator, the completion of a window at operator $o$ triggers a message to be produced at this operator. Window completion is marked by the increment of window ID~\cite{li2008out, li2005semantics}, calculated using the stream's logical time. 
For message $M$ that is sent from upstream operator $o_u$ to downstream operator $o_d$,  $p_{M_{F}}$ can be derived using $p_{M}$ using on a \textsc{Transform} function. With the definition provided by~\cite{li2008out}, \sys defines \textsc{Transform} as:

\vspace{-0.5cm}
\[
\text{$p_{M_{F}}$} =  \textsc{Transform}(p_{M})= \\
\begin{cases}

  \text{$(p_{M}/S_{o_d}+1) \cdot S_{o_d}$} & \text{$S_{o_u} < S_{o_d}$} 
  \\
  \text{$p_{M}$} & \text{otherwise}
\label{eq:transform}
\end{cases}
\]

For a sliding window operator $o_{d}$, $S_{o_d}$ refers to the {\it slide size}, i.e., value step (in terms of logical time) for each window completion to trigger target operator. For the tumbling window operation (i.e., windows cover consecutive, non-overlapping value step), $S_{o_u}$ equals the window size. 
For a message sent by an operator $o_u$ that has a shorter slide size than its targeting operator $o_d$, $p_{M_{F}}$ will be increased to the logical time to trigger $o_d$, that is, $=(p_{M}/S_{o_d}+1) \cdot S_{o_d}$. 

For example if we have a tumbling window with window size 10 s, then the 
expected frontier progress, i.e., $p_{M_{F}}$,
will occur every 10th second (1, 11, 21 ...).
Once the window operator is triggered, the logical time of the resultant message is set to $p_{M_F}$, marking the latest time to  influence a result.%

\noindent\textbf{Step 2 (ProgressMap):} After deriving the frontier progress $p_{M_{F}}$ that triggers the next dataflow output, \sys then estimates the corresponding frontier time $t_{M_{F}}$. 
A temporal data stream typically has its logical time defined in one of three different time domains:\\
(1) {\it event time}~\cite{apache_flink_time,apache_kafka_time}: a totally-ordered value, typically a timestamp, associated with original data being processed; \\
(2) {\it processing time}:  system time for processing each operator~\cite{Akidau2015TheDM};  and\\
(3) {\it ingestion time}: the system time of the data first being observed at the entry point of the system~\cite{apache_flink_time,apache_kafka_time}. \\
\sys  supports both event time and ingestion time. For processing time domain, $M$'s timestamp could be generated when  $M$ is observed by the system.

To generate $t_{M_{F}}$ based on progress $p_{M_{F}}$, \sys utilizes a \textsc{ProgressMap} function to map logical time $p_{M_{F}}$ to physical time $t_{M_{F}}$.
For a dataflow that defines its logical time by data's ingestion time, logical time of each event is defined by the time  when it was observed. Therefore, for {\it all} messages that occur in the resultant dataflow, logical time is assigned by the system at the origin as $t_{M_{F}} = \textsc{ProgressMap}(p_{M_{F}}) = p_{M_{F}}$.

For a dataflow that defines its logical time by the data's event time, $t_{M_{F}} \neq p_{M_{F}}$. %
Our stream processing run-time provides channel-wise guarantee of in-order processing for all target operators. Thus \sys uses linear regression to map $p_{M_{F}}$ to $t_{M_{F}}$, as:
$t_{M_{F}} = \textsc{ProgressMap}(p_{M_{F}}) = \alpha \cdot p_{M_{F}} + \gamma$,
where $\alpha$ and $\gamma$ are parameters derived via a linear fit with running window of historical $p_{M_F}$'s towards their respective $t_{M_F}$'s.
E.g., For same tumbling window with window size 10s, if  $p_{M_{F}}$ occurs at times ($1, 11, 21 \ldots$), with a 2s delay for the event to reach the operator,  $t_{M_{F}}$ will occur at times ($3, 13, 23 \ldots$).

We use a linear model due to our production deployment characteristics:  the data sources %
are largely real time streams, with data ingested soon after generation. 
Users typically expect events to affect results within a constant delay. Thus the logical time (event produced) and the physical time (event observed) are separated by only a small (known) time gap. %
When an event's physical arrival time cannot be inferred from stream progress, we treat windowed operators  as regular operators. Yet, this conservative estimate of  laxity does not hurt performance in practice. %

\section{Scheduling Mechanisms in \sys}
\label{sec:architecture}

\newcommand{\reqContext}{\colorbox{yellow}{1}\xspace}
\newcommand{\reqAPI}{\colorbox{yellow}{3}\xspace}
\newcommand{\reqScheduler}{\colorbox{yellow}{2}\xspace}

We next present \sys's architecture that addresses three main challenges: \\
\reqContext How to make static and dynamic information %
 from both upstream and downstream processing 
 available during priority assignment?\\ %
\reqScheduler How can we efficiently perform fine-grained priority assignment and scheduling that scales with message volume? %
\\
\reqAPI How can we admit pluggable scheduling policies without modifying the scheduler mechanism?

Our approach to address the above challenges is to separate out the priority assignment from scheduling, thus designing a two-level architecture.%
This allows priority assignment for user-defined operators to become %
programmable.%
To pass information between the two levels (and across different operators) we piggyback information atop messages passed between operators.%

More specifically, \sys addresses 
challenge \reqContext
by propagating \textit{scheduling contexts} with messages. 
To meet challenge \reqScheduler,
\sys uses a two-layer scheduler architecture. The top layer, called the \textit{context converter}, is embedded into each operator and 
handles scheduling contexts
whenever the operator sends or receives a message.
The bottom layer, called the \textit{\sys scheduler}, interprets message priority based on the scheduling context embedded within a message and updates a priority-based data structure for both operators and operators' messages. %
Our design has advantages of: 
(i) avoiding the bottleneck of having a centralized scheduler thread calculate priority for each operator upon arrival of messages, and (ii) only limiting priority to be per-message. %
This  allow the operators, dataflows, and the scheduler, to all remain  stateless. 

To address \reqAPI \sys allows the priority generation process
to be implemented through the context handling API. %
A context converter invokes the API with each operator.%

\subsection{Scheduling Contexts}
\label{subsec:contexts}
Scheduling contexts are data structures attached to messages, capturing message priority, {and} information required to perform priority-based scheduling. Scheduling contexts are \textit{created}, \textit{modified}, and \textit{relayed} alongside their respective messages. 
Concretely, scheduling contexts allow capture of  scheduling states
of both upstream and downstream execution.  
A scheduling context can be seen and modified by %
{both} context converters and the \sys scheduler. 
There are two kinds of contexts: 

1. 
\textbf{Priority Context} (\texttt{PC}): \texttt{PC} is necessary for the scheduler to infer the priority of a message.
In \sys \texttt{PC}s are defined to include local and global priority as ($ID$, $PRI_{local}$, $PRI_{global}$, $Dataflow\_DefinedField$). 
$PRI_{local}$ and $PRI_{global}$ are used for applications to enclose message priorities for scheduler to determine execution order, and $Dataflow\_DefinedField$ includes upstream information required by the pluggable policy to generate message priority. 

A \texttt{PC} is attached to a message before the message is sent. It is either created at a source operator upon receipt of an event, or inherited and modified from the upstream message that triggers the current operator. Therefore, a \texttt{PC} is seen and modified by all executions of upstream operators that lead to the current message. This enables \texttt{PC} to address challenge~\reqContext by capturing information of dependant upstream execution (e.g., stream progress, latency target, etc.).

2. 
\textbf{Reply Context} (\texttt{RC}): \texttt{RC} meets challenge~\reqContext by capturing periodic feedback from the downstream operators. 
\texttt{RC} is attached to an acknowledgement message 
~\footnote{A common approach used by many stream processing systems~\cite{storm, heron, flink} to ensure processing correctness},
sent by the target operator to its upstream operator after a message is received. \texttt{RC}s provide processing feedback of the target operator and all its downstream operators. \texttt{RC}s can be aggregated and relayed recursively upstream through the dataflow.%

\sys provides a programmable API to implement these scheduling contexts and their corresponding policy handlers in context converters.
API functions include: 

1. \textbf{function}~\textsc{BuildCxtAtSource(Event $e$)} that creates a \texttt{PC} upon receipt of an event $e$; 

2. \textbf{function}~\textsc{BuildCxtAtOperator(Message $M$)} that modifies and propagates a \textsc{PC} when an operator is invoked (by $M$) and ready to send a message downstream; %

3. \textbf{function}~\textsc{ProcessCtxFromReply(Message $r$)} that processes \textsc{RC} attached to an acknowledgement message $r$ {received at  upstream operator}; and 
 
4. \textbf{function}~\textsc{PrepareReply(Message $r$)} that generates \textsc{RC} containing user-defined feedbacks, attached to $r$ sent by a downstream operator.%
 
\subsection{System Architecture}
\label{subsec:architecture}

\begin{figure}[!bt]
\centering
\begin{tabular}{c}
\subfigure[
\textit{Scheduling contexts circulating between two operators.}
]{	
	\includegraphics[width=8.5cm]{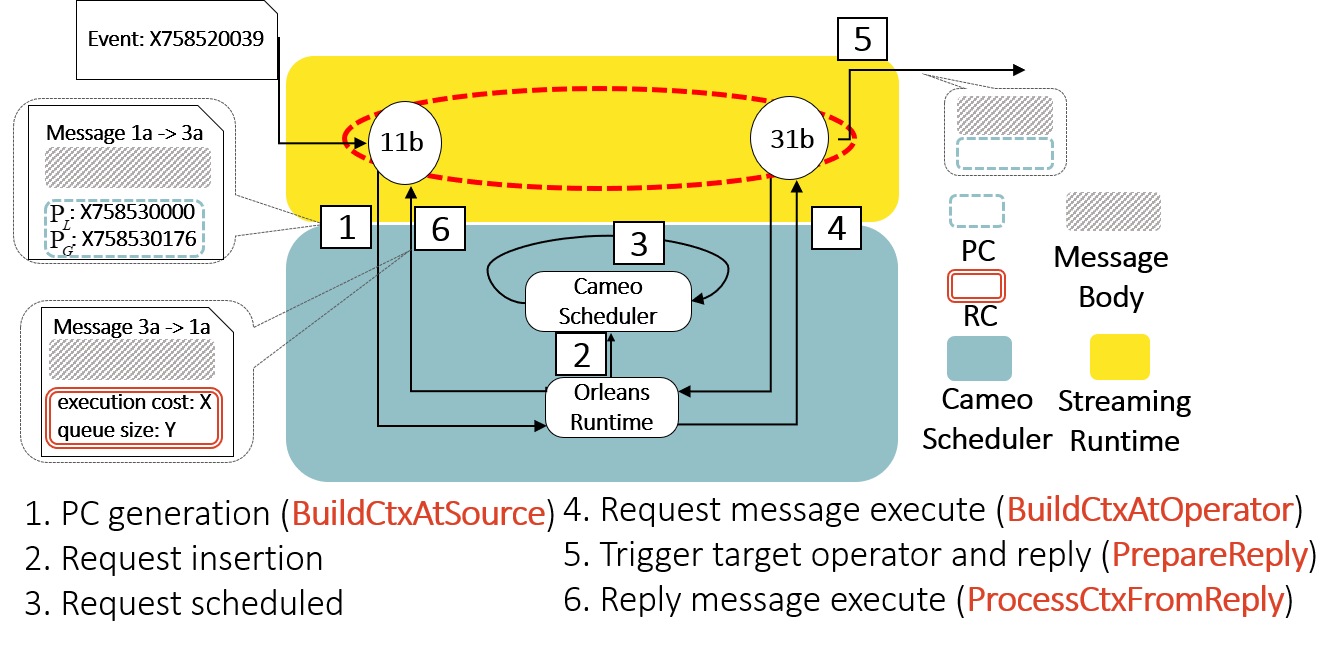}
	\label{fig:converter}	
}
\vspace{-0.2cm}
\\
\subfigure[\textit{\sys Scheduler Architecture. Operators sorted by global priority. Messages at an operator sorted by local priority.}]{
	\includegraphics[width=8.5cm]{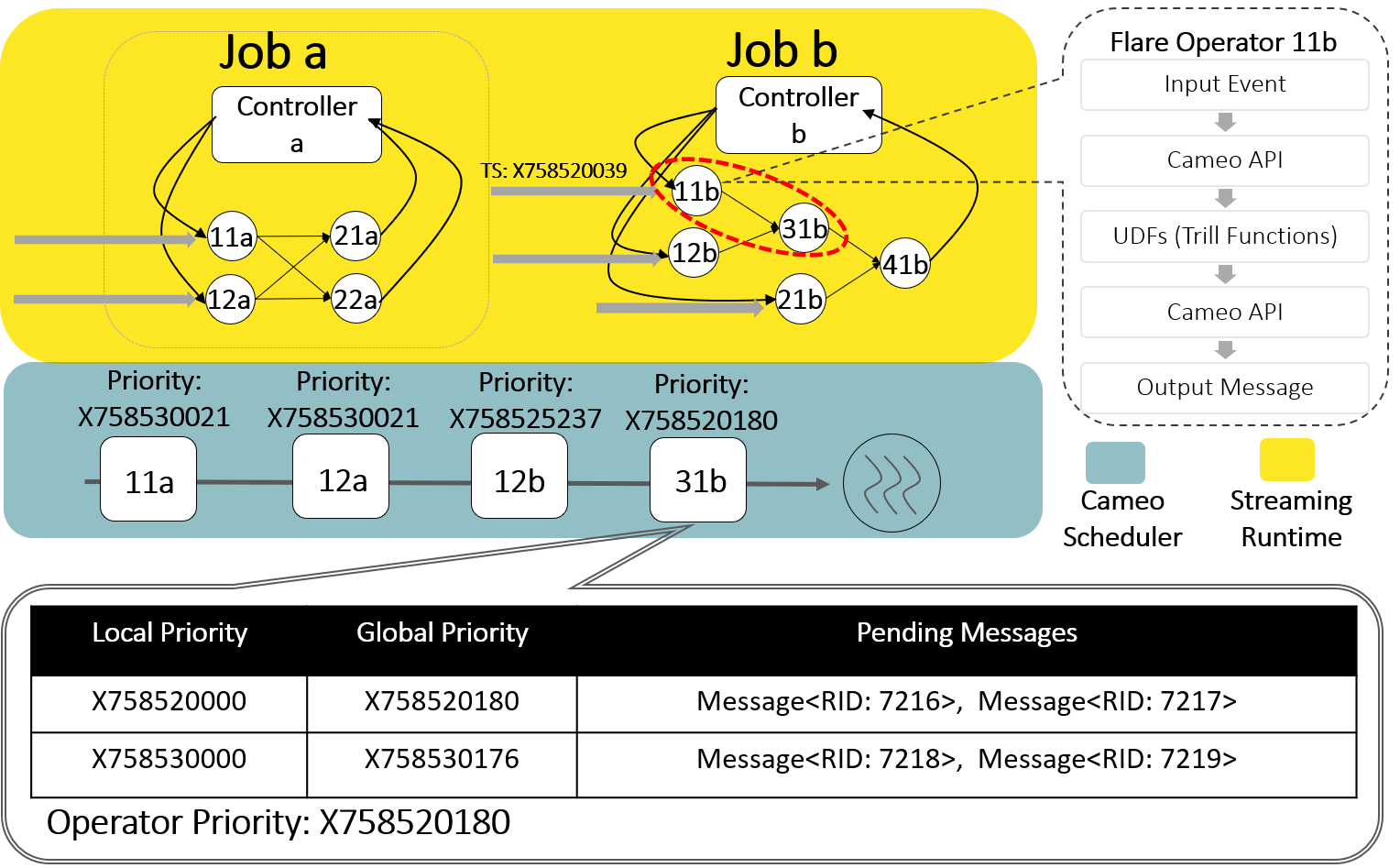}
	\label{fig:scheduler}
}

\end{tabular}
\vspace{-0.2cm}
\caption{
\textit{
\sys Mechanisms.
}
}
\vspace{-0.4cm}
\label{fig:architecture}
\end{figure}

Figure~\ref{fig:converter} shows context converters at work.
After an event is generated at a source operator $1a$ (step 1), the converter creates a \texttt{PC} through \textsc{BuildCxtAtSource} and sends the message to \sys scheduler. The target operator is scheduled (step 2) with the priority extracted from the \texttt{PC}, before it is executed. Once the target operator $3a$ is triggered (step 4), it calls \textsc{BuildCtxAtOperator}, modifying and relaying $PC$ with its message to downstream operators.
After that $3a$ sends an acknowledgement message with an \texttt{RC} (through \textsc{PrepareReply})
back to $1a$ (step 5). 
\texttt{RC} is then populated by the scheduler with  runtime statistics (e.g, CPU time, queuing delays, message queue sizes, network transfer
time, etc.) before it is scheduled and delivered at the source operator (step 6).

\sys enables scheduling states to be managed and transported alongside the data.   
This allows \sys to meet challenge \reqScheduler by keeping the scheduler away from centralized state maintenance and priority generation. 
The \sys scheduler manages a two level priority-based data structure, shown in Figure~\ref{fig:scheduler}. We use $PRI_{local}$ to determine $M$'s execution priority within its target operator, and  $PRI_{global}$ of the next message in an operator to order all operators that have pending messages. 
\sys can schedule at either message granularity or a
coarser time quanta. While processing a message, \sys \textit{peeks} at the priority of the next operator in the  queue. If the next operator has higher priority, we swap with the current operator after a fixed time quantum (tunable).

\subsection{Implementing the \sys Policy}
\label{subsec:context-conversion}

To implement the scheduling policy of  Section~\ref{sec:scheduling}, a \texttt{PC} is attached to message $M$ (denoted as $PC(M)$) with these fields: %

\vspace{.2cm}
\resizebox{.38\textwidth}{!}{
\begin{minipage}{.45\textwidth}
\begin{center}
\begin{tabular}[h!]{ |c|c|c|c| } 
\hline
$ID$ & $PRI_{local}$ & $PRI_{global}$ & $Dataflow-DefinedField$\\
\hline
$ID_{M}$ & $p_{M_{F}}$ & $ddl_{M_{F}}$ & $(p_{M_{F}}, t_{M_{F}}, L)$ \\
\hline
\end{tabular}
\end{center}
\end{minipage} }

\begin{algorithm}[tb!]
\caption{Priority Context Conversion}
\label{alg:converter}
\begin{algorithmic}[1]
\algnewcommand{\IIf}[1]{\State\algorithmicif\ #1\ \algorithmicthen}
\algnewcommand{\IElse}[1]{\State\algorithmicelse\ #1\ }
\algnewcommand{\EndIIf}{\unskip\ \algorithmicend\ \algorithmicif}	

\Function{BuildCxtAtSource}{\textsc{Event} $e$} \Comment {Generate {\tt PC} for message $M_{e}$ at source triggered by event $e$}
\State ${\tt PC}(M_{e}) \leftarrow \Call{InitializePriorityContext}$
\State ${\tt PC}(M_{e}).(PRI_{local},PRI_{global}) \leftarrow (e.p_{e},e.t_{e})$
\State ${\tt PC}(M_{e}) \leftarrow \Call{ContextConvert}{{\tt PC}(M_{e}), {\tt RC_{local}}}$

\State \Return ${\tt PC}(M_{e})$

\EndFunction

\Function{BuildCxtAtOperator}{\textsc{Message} $M_{n}$} \Comment {Generate {\tt PC} for message $M_d$ at an intermediate operator triggered by upstream message $M_u$}
\State ${\tt PC}(M_d) \leftarrow {\tt PC}(M_u)$
\State ${\tt PC}(M_d).(PRI_{local}, PRI_{global})\leftarrow {\tt PC}(M_u).(p_{M_F}, t_{M_F})$
\State ${\tt PC}(M_d) \leftarrow \Call{ContextConvert}{{\tt PC}(M_d), {\tt RC_{local}}}$

\State \Return ${\tt PC}(M_d)$

\EndFunction

\Function{CxtConvert}{${\tt PC}(M), {\tt RC}$}\Comment {Calculating message priority based on ${\tt PC}(M), {\tt RC}$ provided}
\State ${p_{M_{F}} \leftarrow \Call{Transform}{{\tt PC}(M).p_{M}}}$
\State $t_{M_{F}} \leftarrow \Call{ProgressMap}{p_{M_{F}}}$ \Comment{As in Section~\ref{subsec:propagation}}

\If {$t_{M_F}$ defined in stream event time}
	\State $\Call{ProgressMap.update}{{\tt PC}.t_{M}, {\tt PC}.p_{M}}$ \Comment{Improving prediction model as in Section~\ref{subsec:context-conversion}
	}
\EndIf

\State ${\tt PC}(M).p_{M}, {\tt PC}(M).t_{M} \leftarrow p_{M_{F}}, t_{M_{F}}$

\State $ddl_{M} \leftarrow t_{M_{F}} + {\tt PC}(M).L - {\tt RC}.C_{m}-{\tt RC}.C_{path}$
\State ${\tt PC}(M).(PRI_{local},PRI_{global}) \leftarrow (p_{M_F}, ddl_{M})$

\EndFunction

\Function{ProcessCtxFromReply}{\textsc{Message} $r$} \Comment {Retrieve reply message's ${\tt RC}$ and store locally}
\State ${\tt RC_{local}}.update(r.{\tt RC})$ 

\EndFunction

\Function{PrepareReply}{\textsc{Message} $r$} \Comment Recursively update maximum critical path cost $C_{path}$ before reply
\If {\textsc{Sender}$(r)$ = Sink} 
\State $r.{\tt RC} \leftarrow \Call{InitializeReplyContext}$
\Else {} $r.{\tt RC}.C_{path} \leftarrow {\tt RC}.C_{m}+{\tt RC}.C_{path}$
\EndIf
\EndFunction

\end{algorithmic}
\end{algorithm}

The core of Algorithm~\ref{alg:converter} is \textsc{CxtConvert}, which generates \texttt{PC} for downstream message $M_d$ (denoted as \texttt{PC}$(M_d)$), triggered by \texttt{PC}$(M_u)$ from the upstream triggering message.
To schedule a downstream message $M_d$ triggered by $M_u$, \sys first retrieves stream progress $p_{M_u}$ contained in \texttt{PC}$(M_u)$. 
It then applies the two-step process %
(Section~\ref{subsec:propagation}) to calculate frontier time $t_{M_{F}}$ using $p_{M_u}$.
This  may extend a message's deadline if the operator is not expected to trigger immediately (e.g., windowed operator). 
We capture $p_{M_{F}}$ and %
estimated $t_{M_{F}}$ in %
\texttt{PC} as message priority and propagate this downstream.   
Meanwhile,  $p_{M_u}$ and $t_{M_u}$
are fed into a linear model to improve future prediction towards $t_{M_{F}}$. Finally, the context converter computes message priority $ddl_{M_u}$ using $t_{M_{F}}$ as described in Section~\ref{sec:scheduling}.

\sys utilizes \texttt{RC} to  track  critical path execution cost $C_{path}$ and execution cost $C_{o_M}$.
\texttt{RC} contains the processing cost (e.g., CPU time) of the downstream critical path up to the current operator, obtained via profiling.

\subsection{Customizing \sys: Proportional Fair Scheduling}
\label{subsec:discussion}

We next show how the pluggable scheduling policy in \sys can be used to support other performance objectives, thus satisfying \reqAPI.
For instance, we show how a token-based rate control mechanism works, where token rate %
{equals} desired output rate. 
In this setting, each application is granted tokens per unit of time, based on their target sending rate. If a source operator exceeds its target sending rate, 
the remaining messages (and all downstream traffic) are processed 
with operator priority reduced to minimum. 
When capacity is insufficient to meet the aggregate token rate, all dataflows are downgraded equally. 
\sys spreads tokens proportionally across the next time interval (e.g., 1 sec) by tagging each token with the timestamp at each source operator. For token-ed messages, we use token tag  $PRI_{global}$, and interval ID as $PRI_{local}$. Messages without tokens have $PRI_{global}$ set to \texttt{MIN\_VALUE}. Through \texttt{PC} propagation, all downstream messages are processed when no tokened traffic is present.

\begin{figure}
\centering
\includegraphics[width=.42\textwidth]{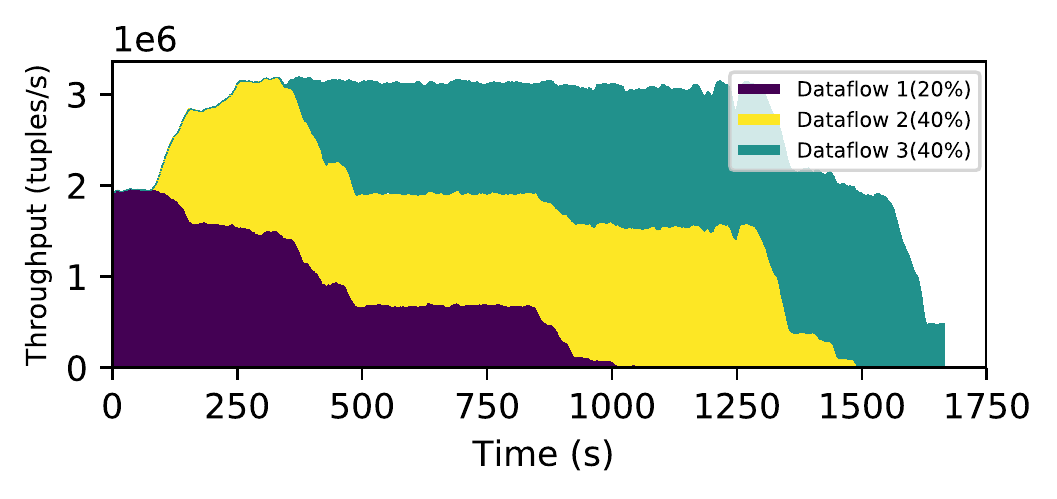}
\vspace{-0.1in}
\caption{
\textit{Proportional fair sharing using \sys. %
}
}
\vspace{-.2in}
\label{fig:example4}
\end{figure}

Figure~\ref{fig:example4} shows \sys's token mechanism. Three dataflows start with 20\% (12), 40\% (24), and 40\% (24) tokens as target ingestion rate per source respectively. Each ingests 2M events/s, starting 300 s apart, and lasting 1500 s. Dataflow 1 receives full capacity initially when there is no competition. The cluster is at capacity after Dataflow 3 arrives, but \sys ensures token allocation translates into throughput shares. %

\section{Experimental Evaluation}
\label{sec:evaluation}

We next present experimental evaluation of \sys. We first study the effect of different queries on \sys in a single-tenant setting. Then for multi-tenant settings, we study \sys's effects when: 

\squishlist
    \item \textbf{Varying environmental parameters} (Section~\ref{subsec:workload_adaptability}): 
    This includes: a) workload (tenant sizes and ingestion rate), and b) available resources, i.e., worker thread pool size, c) workload bursts.
    \item \textbf{Tuning internal parameters and optimization} (Section~\ref{subsec:internal_variation}):
    We study: a) effect of scheduling granularity, b) frontier prediction for event time windows, and c) starvation prevention.
\squishend

We implement streaming queries in Flare~\cite{mai2018chi} (built atop Orleans~\cite{bykov2011orleans, orleans_dotnet}) by using Trill~\cite{chandramouli2014trill}  to run streaming operators.%
We compare \sys vs. both i) default {\bf Orleans} (version 1.5.2) scheduler, and ii) a custom-built {\bf FIFO} scheduler. %
By default, we use the 1 ms minimum re-scheduling grain  (Section~\ref{subsec:architecture}). This grain is generally shorter than a message's execution time. %
Default Orleans implements a global run queue of messages using a ConcurrentBag~\cite{concurrentbag} data structure. ConcurrentBag optimizes processing throughput by prioritizing processing thread-local tasks over the global ones. 
For the FIFO scheduler, we insert operators into the global run queue and extract them in FIFO order. In both approaches, an operator processes its messages in FIFO order. %

\label{subsec:eval_setting}

\mypar{Machine configuration} We use DS12-v2 Azure virtual machines (4 vCPUs/56GB memory/112G SSD) as server machines, and DS11-v2 Azure virtual machines (2 vCPUs/14GB memory/28G SSD) as  client machines \cite{azuresizes}. Single-tenant scenarios are evaluated on a single server machine. Unless otherwise specified, all multi-tenant experiments are evaluated using a 32-node Azure cluster with 16 client machines.

\mypar{Evaluation workload}
For the multi-job setting we study performance isolation under concurrent dataflow jobs. Concretely, our workload is divided into two control groups:

\squishlist
\item {\bf Latency Sensitive Jobs (Group 1 ):} This is representative of jobs connected to user dashboards, or associated with SLAs, ongoing advertisement campaigns, etc. Our workload jobs in Group 1  have sparse input volume across time (1 msg/s per source, with 1000 events/msg), and report periodic results with shorter aggregation windows (1 second). These have strict latency constraints.

\item {\bf Bulk Analytic Jobs (Group 2):} This is representative of social media streams being processed into longer-term analytics with longer aggregation windows (10 seconds). Our Group 2 jobs have input of both higher and variable volume and high speed, but with lax latency constraints.
\squishend

Our queries feature multiple stages of windowed aggregation parallelized into a group of operators. 
Each job has 64 client sources. 
All queries assume input streams associated with event time unless specified otherwise.

\mypar{Latency constraints} In order to determine the  latency constraint of one job, we run multiple instances of the job until the resource (CPU) usage reaches 50\%. Then we set the  latency constraint of the job to be twice the tail (95th percentile) latency. This emulates the scenario where users with experience
in isolated environments deploy the same query 
in a shared environment by moderately relaxing the latency constraint. Unless otherwise specified, a latency target is marked with grey dotted line in the plots.

\subsection{Single-tenant Scenario}
\label{subsubsec:single_job}
\begin{figure}[!bt]
\centering
 \setlength\extrarowheight{-0.4in}
\begin{tabular}{cc}
\subfigure[%
]{
  \includegraphics[height=1.2in]{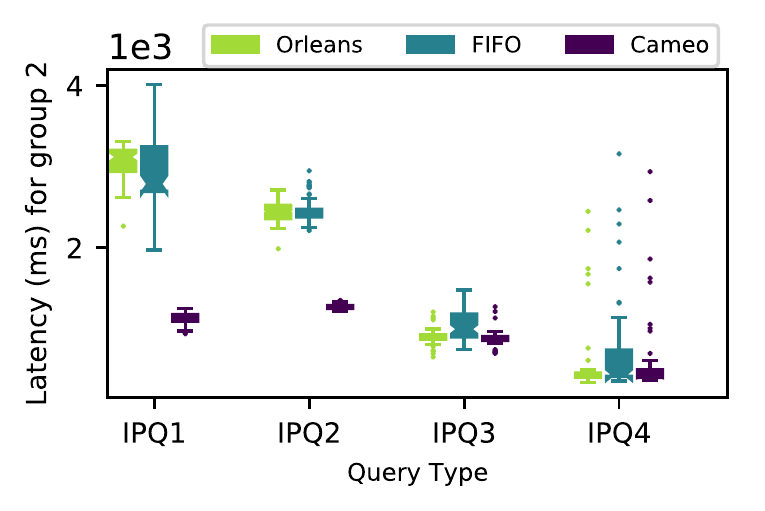}
\label{exp:single_query_boxes}
\vspace{-0.2in}
}
\hspace{-.25in}
&
\subfigure[%
]{
  \includegraphics[height=1.2in]{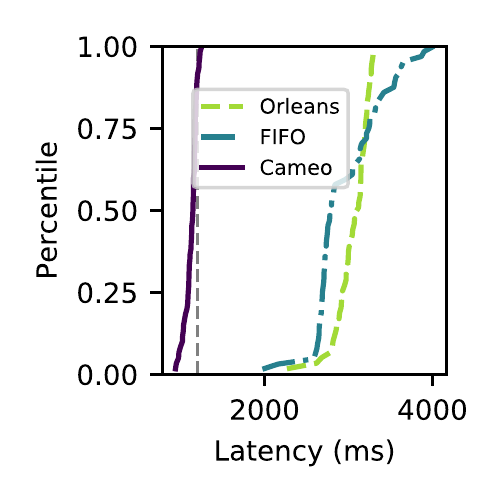}
\label{exp:single_query_cdf}
\vspace{-0.2in}
}
\\
\multicolumn{2}{c}{
\vspace{-0.2cm}
\subfigure[
]
{
\vspace{-0.1in}
\includegraphics[width=0.46\textwidth]{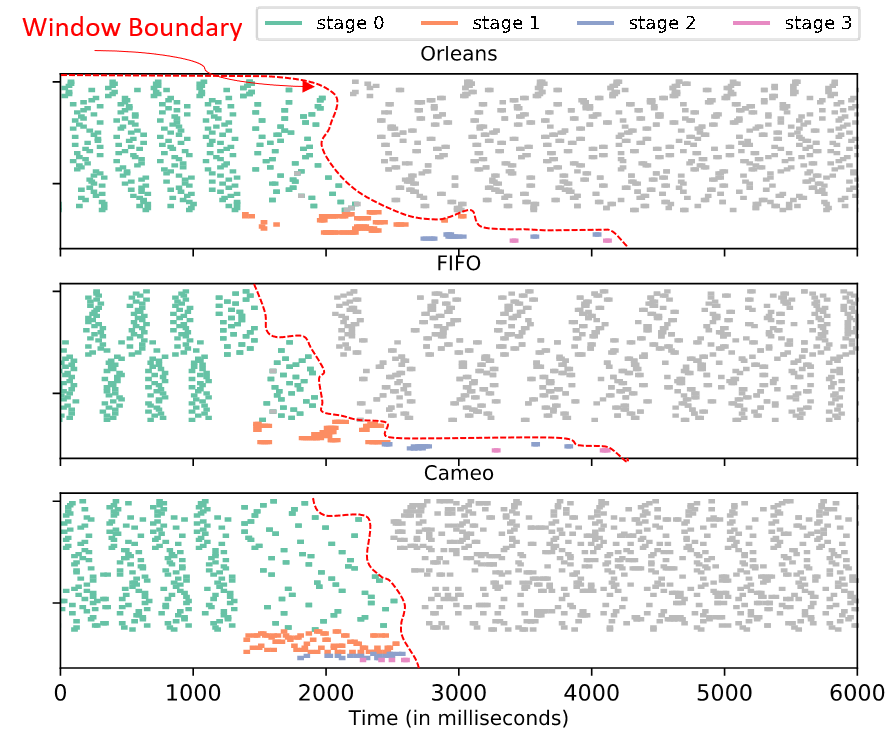}
\label{exp:single_timelines}
} 
}
\end{tabular}
\caption{
\it Single-Tenant Experiments: (a) Query Latency. %
(b) Latency CDF. (c) Operator Schedule Timeline: X axis = time when operator was scheduled. Y axis = operator ID color coded by operator's stage. 
Operators are triggered at each stage in order (stage 0 to 3). Job latency is time from all events that belong to the previous window being received at stage 0, until last message is output at stage 3. %
}
\vspace{-.2in}
\label{exp:scheduling_single_dtflw}
\end{figure}

In Figure~\ref{exp:scheduling_single_dtflw} we evaluate a single-tenant setting with 4 queries: IPQ1 through IPQ4. IPQ1 and IPQ3 are periodic and they respectively calculate sum of revenue generated by real time ads, and the number of events generated by jobs groups by different criteria. IPQ2 performs similar aggregation operations as IPQ1 but on a sliding window (i.e., consecutive window contains overlapped input). IPQ4 summarizes errors from log events via running a windowed join of two event stream, followed by aggregation on a tumbling window (i.e., where consecutive windows contain non-overlapping ranges of data that are evenly spaced across time).

From Figure~\ref{exp:single_query_boxes} 
we observe that \sys improves 
median latency by up to %
2.7$\times$%
and tail latency by up to %
3.2$\times$.%
We also observe that default Orleans performs almost as well as \sys for  IPQ4.
This is because IPQ4 has a higher execution time with heavy memory access, and performs well when pinned to a single thread with better access locality.

\mypar{Effect on intra-query operator scheduling} 
The CDF in Figure~\ref{exp:single_query_cdf} shows %
that Orleans' latency is about 3$\times$ higher than \sys. While FIFO has a slightly lower median latency, its tail latency is as high as in Orleans.

\sys's prioritization is especially evident in  Figure~\ref{exp:single_timelines}, where dots are message starts, and red lines separate windows. We first observe that \sys is faster, and it creates a clearer boundary between windows. Second, messages that contribute to the first result (colored dots) and messages that contribute to the second result (grey dots) do not overlap on the timeline. For the other two strategies, there is a {\it drift} between stream progress in early stages vs. later stages, producing a prolonged delay. In particular, in Orleans and FIFO, early-arriving messages from the next window are executed {\it before} messages from the first window, thus missing deadlines.

\subsection{Multi-tenant Scenario}
\label{subsec:workload_adaptability}

\begin{figure} [t]
\centering
\vspace{-0.1in}

\begin{tabular}{c}

\subfigure[{\it Varying ingestion rate of group 2 tenants (Bulk Analytics). }]{
\hspace{-0.05in}
    \includegraphics[width=0.48\textwidth]{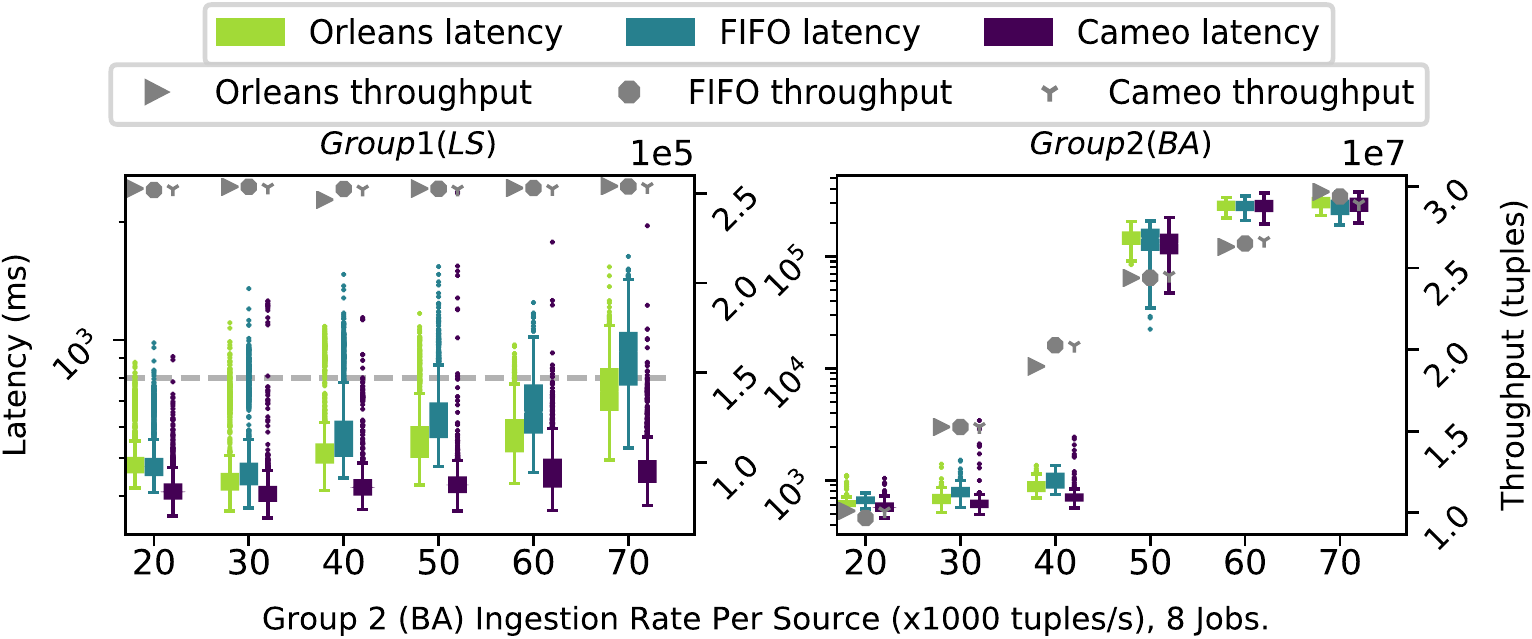}
    \label{exp:dataflow_ingestion_rate}
}
\\

\subfigure[{\it 
Varying number of group 2 tenants (Bulk Analytics). }]{
\hspace{-0.05in}
    \includegraphics[width=0.48\textwidth]{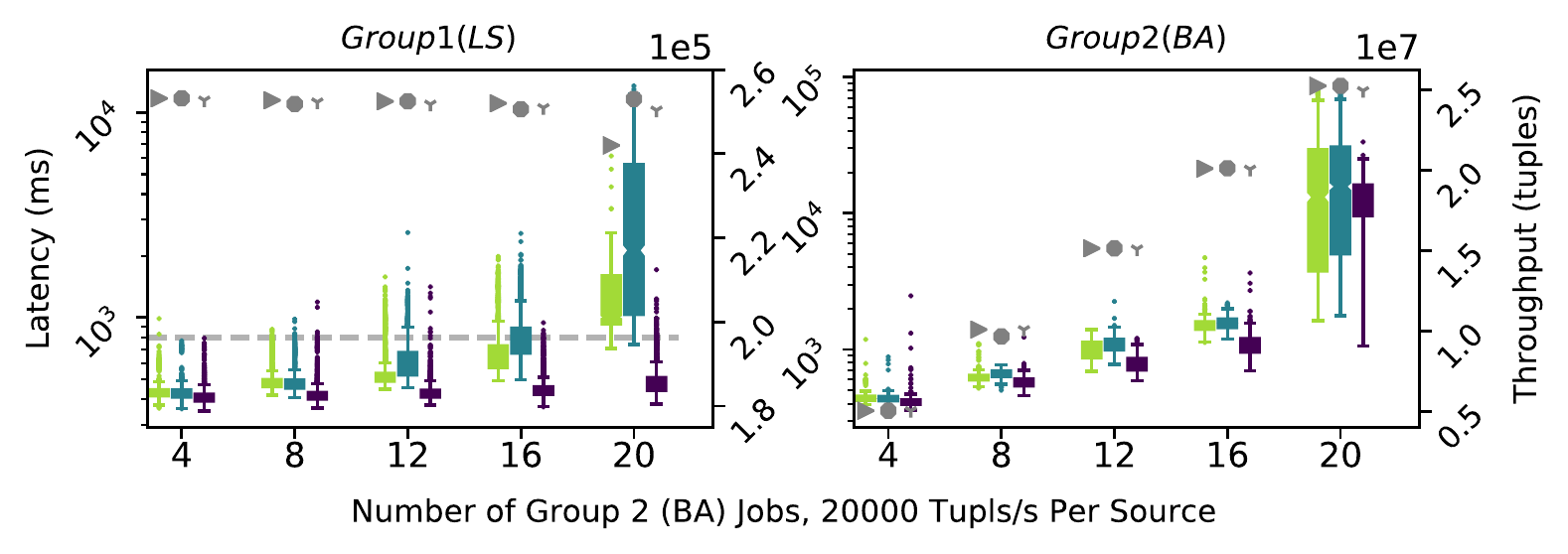}
    \label{exp:dataflow_count}
}
\\
\vspace{-0.05in}

\subfigure[{\it 
 Varying worker thread pool size.}]{
 \hspace{-0.05in}
    \includegraphics[width=0.48\textwidth]{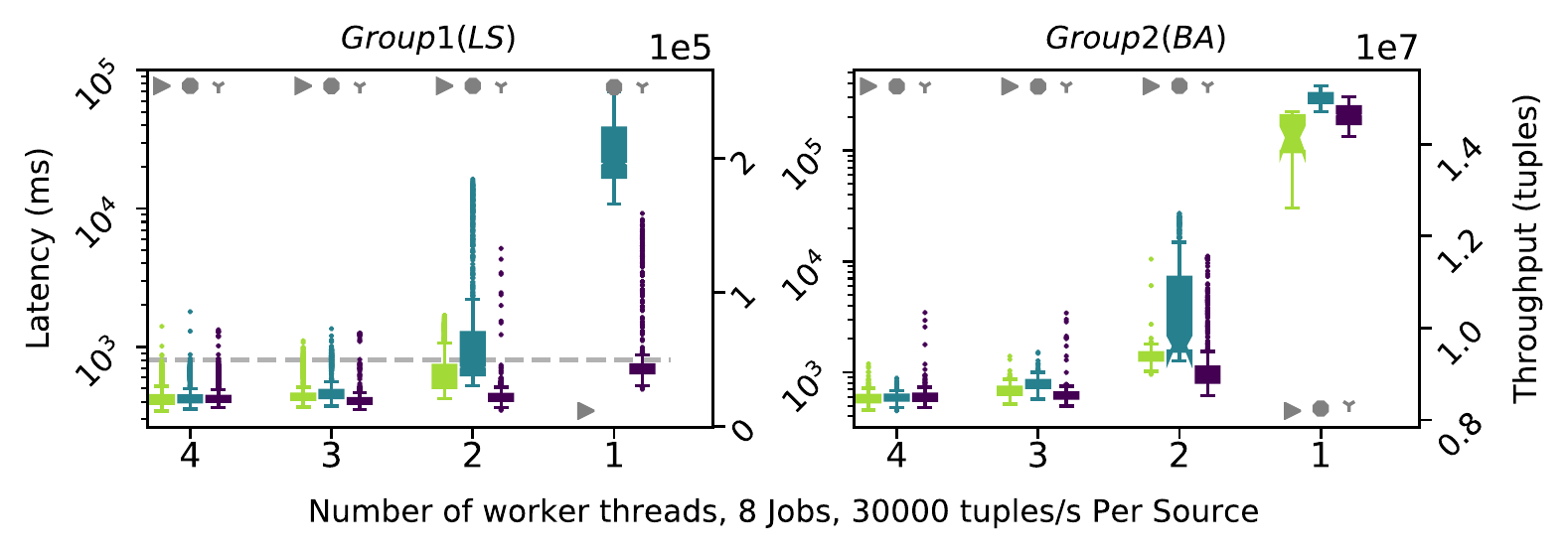}
    \label{exp:worker_count}
}
\end{tabular}
\vspace{-0.05in}
\caption{\it 
Latency-sensitive jobs under %
competing workloads. %
}
\label{LSexpt}
\vspace{-0.2in}
\end{figure}

Figure~\ref{LSexpt} studies a control group of latency-constrained dataflows (group 1 LS jobs) by fixing both job count and data ingestion rate. %
We vary data volume from competing workloads (group 2 BA jobs) and available resources. For LS jobs we impose a latency target of 800 ms, while for BA jobs we use a 7200s latency constraint. %

\noindent \textbf{\sys under increasing data volume.}
We run four group 1 jobs alongside group 2 jobs. We increase the competing group 2 jobs' traffic, by increasing the ingestion speed (Figure~\ref{exp:dataflow_ingestion_rate}) and number of tenants (Figure~\ref{exp:dataflow_count}).
We observe that all three strategies (\sys, Orleans, FIFO) are comparable up to per-source tuple rate of %
30K/s %
in Figure~\ref{exp:dataflow_ingestion_rate}, and up to {twelve} group 2 jobs in Figure~\ref{exp:dataflow_count}.
Beyond this, overloading causes massive latency degradation, for group 1 (LS) jobs at median and 99 percentile latency (respectively): i) Orleans %
is worse than \sys by %
up to 1.6 and 1.5$\times$ in Figure~\ref{exp:dataflow_ingestion_rate}, up to 2.2 and 2.8$\times$ in Figure~\ref{exp:dataflow_count},  and ii) FIFO %
is worse than \sys by %
up to 2 and 1.8$\times$ in Figure~\ref{exp:dataflow_ingestion_rate}, up to 4.6 and 13.6$\times$ in Figure~\ref{exp:dataflow_count}.
\sys stays stable. %
\sys's degradation of group 2 jobs is small--- with latency similar or lower than Orleans and FIFO, and \sys's throughput only 2.5\% lower.%

\noindent \textbf{Effect of limited resources.} 
Orleans'~\cite{newell2016optimizing} underlying SEDA architecture~\cite{welsh2001seda}
resizes thread pools to achieve resource balance between execution steps, for dynamic re-provisioning. %
Figure~\ref{exp:worker_count} shows latency and throughput when we {decrease} the number of worker threads. \sys maintains the performance of group 1 jobs except in the most restrictive 1 thread case (although it still meets %
90\% %
of deadlines). 
\sys prefers messages with impending deadlines and this causes back-pressure for jobs with less-restrictive latency constraints, lowering  throughput.
Both Orleans and FIFO observe large performance penalties for group 1 and 2 jobs (higher in the former). Group 2 jobs with much higher ingestion rate will naturally receive more resources upon message arrivals, leading to back-pressure  and lower throughput for group 1 jobs.

\begin{figure}[!bt]
\centering
\begin{tabular}{cc}

\hspace{-.2in}
\subfigure[{\it Orleans Latency Timeline}]{
  \includegraphics[height=1.3in]{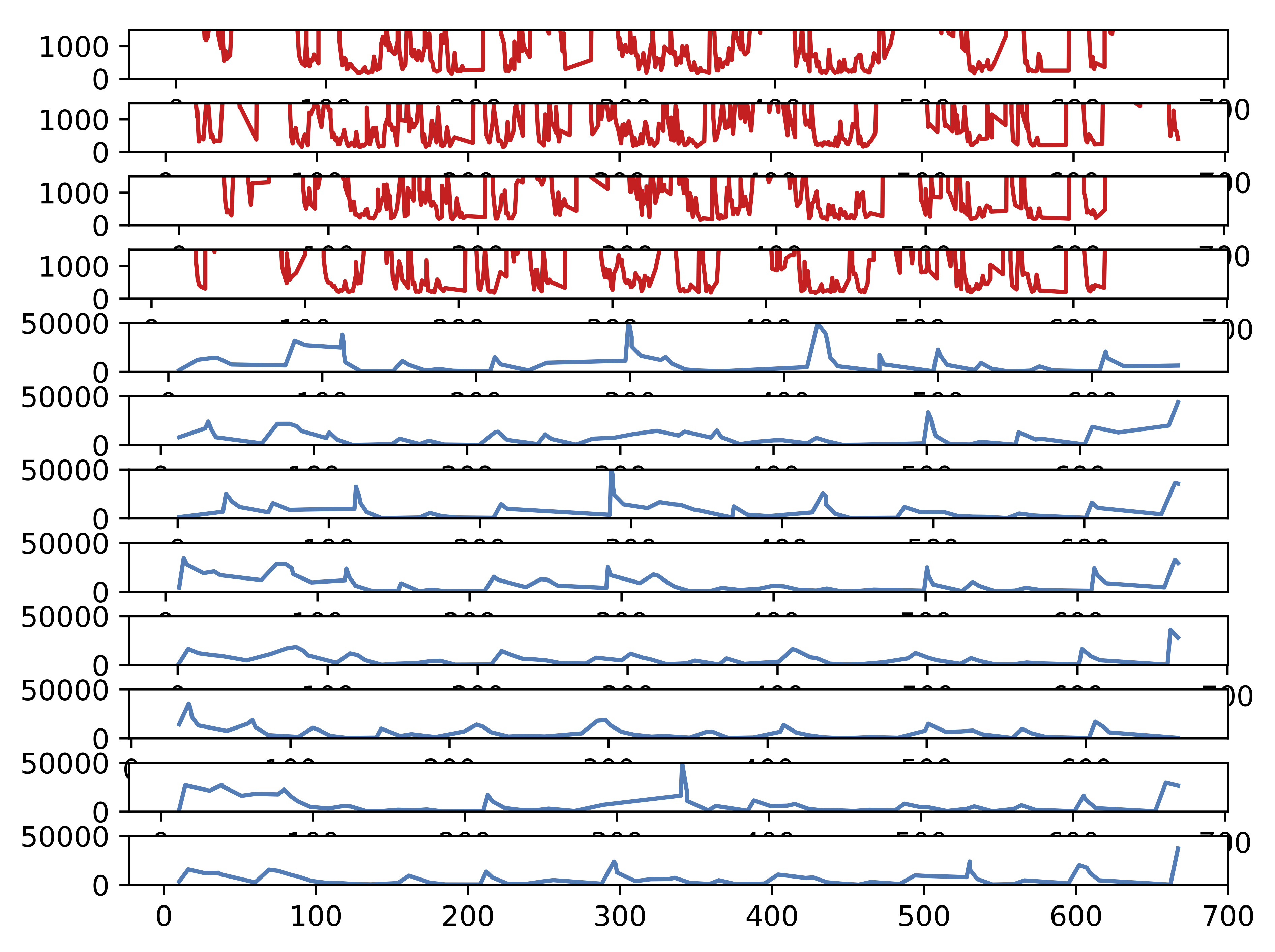}
\label{exp:orleans_pareto_timeline}
}
\hspace{-.3in}
& 
\subfigure[{\it FIFO Latency Timeline}]{
  \includegraphics[height=1.3in]{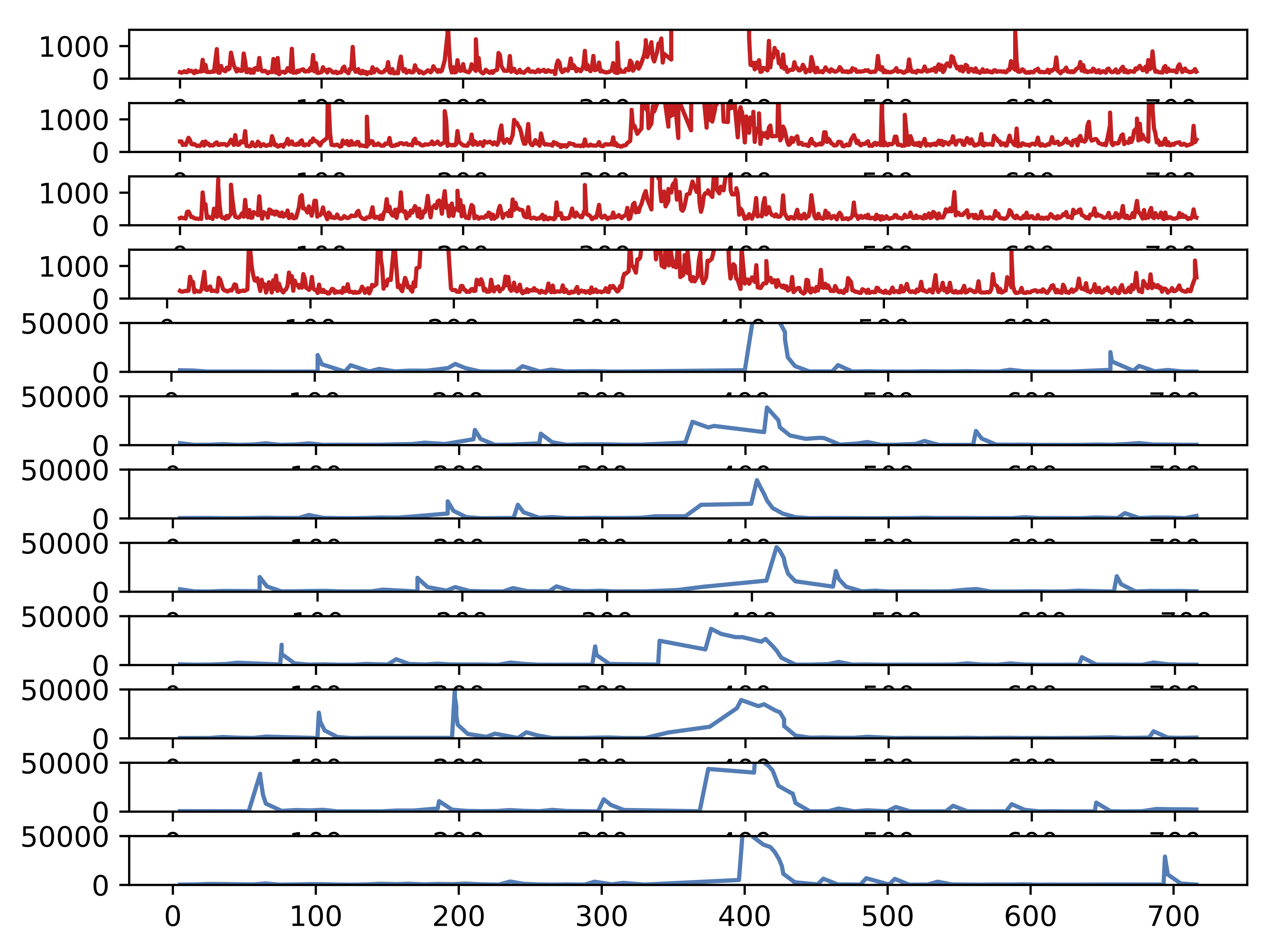}
\label{exp:fifo_pareto_timeline}
}
\vspace{-.1in}
 \\
\hspace{-.2in}
\subfigure[{\it Cameo Latency Timeline}]{
  \includegraphics[height=1.3in]{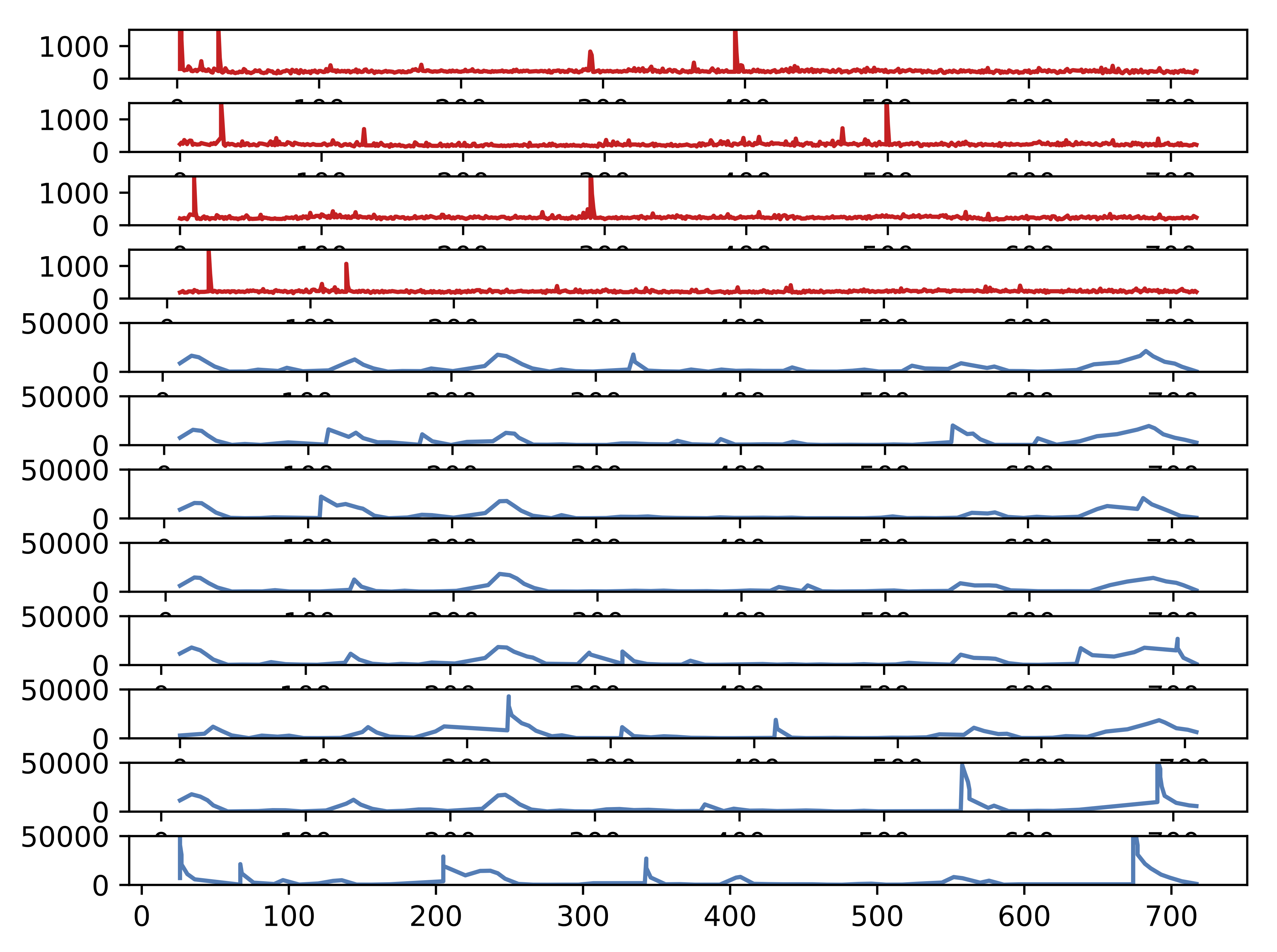}
\label{exp:cameo_pareto_timeline}
}
\hspace{-.3in}
& 
\subfigure[{\it Latency Distribution}]{
  \includegraphics[height=1.25in]{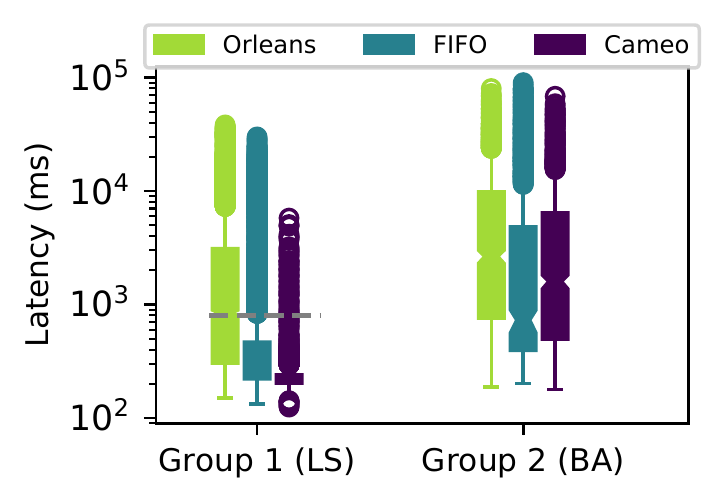}
\label{exp:pareto_boxes}
}

\end{tabular}
\vspace{-0.1in}
\caption{\it %
Latency  under Pareto event arrival. 
}
\vspace{-.1in}
\label{exp:pareto_mixes}
\end{figure}

\noindent \textbf{Effect of temporal variation of workload.} 
We use a Pareto distribution for data volume  in Figure~\ref{exp:pareto_mixes}, with four group 1 jobs and eight group 2 jobs. (This is based on  Figures~\ref{fig:volume_distribution},  \ref{fig:volume_heatmap}, which showed a Power-Law-like distribution.) The cluster utilization 
is kept under 50\%.  

High ingestion rate can suddenly prolong queues at machines. Visualizing  timelines in  Figures~\ref{exp:orleans_pareto_timeline}, \ref{exp:fifo_pareto_timeline}, and \ref{exp:cameo_pareto_timeline} shows that for latency-constrained jobs (group 1), \sys's latency is more stable than Orleans' and FIFO's. Figure~\ref{exp:pareto_boxes} shows that \sys %
reduces (median, 99th percentile) latency by (3.9$\times$, 29.7$\times$) vs. Orleans, and (1.3$\times$, 21.1$\times$) vs. FIFO. %
\sys's standard deviation is also lower by 
23.2$\times$ %
and %
12.7$\times$ %
compared to Orleans and FIFO respectively. 
For group 2, \sys produces smaller average latency and is less affected by ingestion spikes. Transient workload bursts affect many jobs, e.g., %
all jobs %
around %
$t=400$ %
with 
FIFO, %
as a spike at one operator affects all its collocated operators.  %

\begin{figure}[!bt]
\centering

\includegraphics[width=0.5\textwidth]{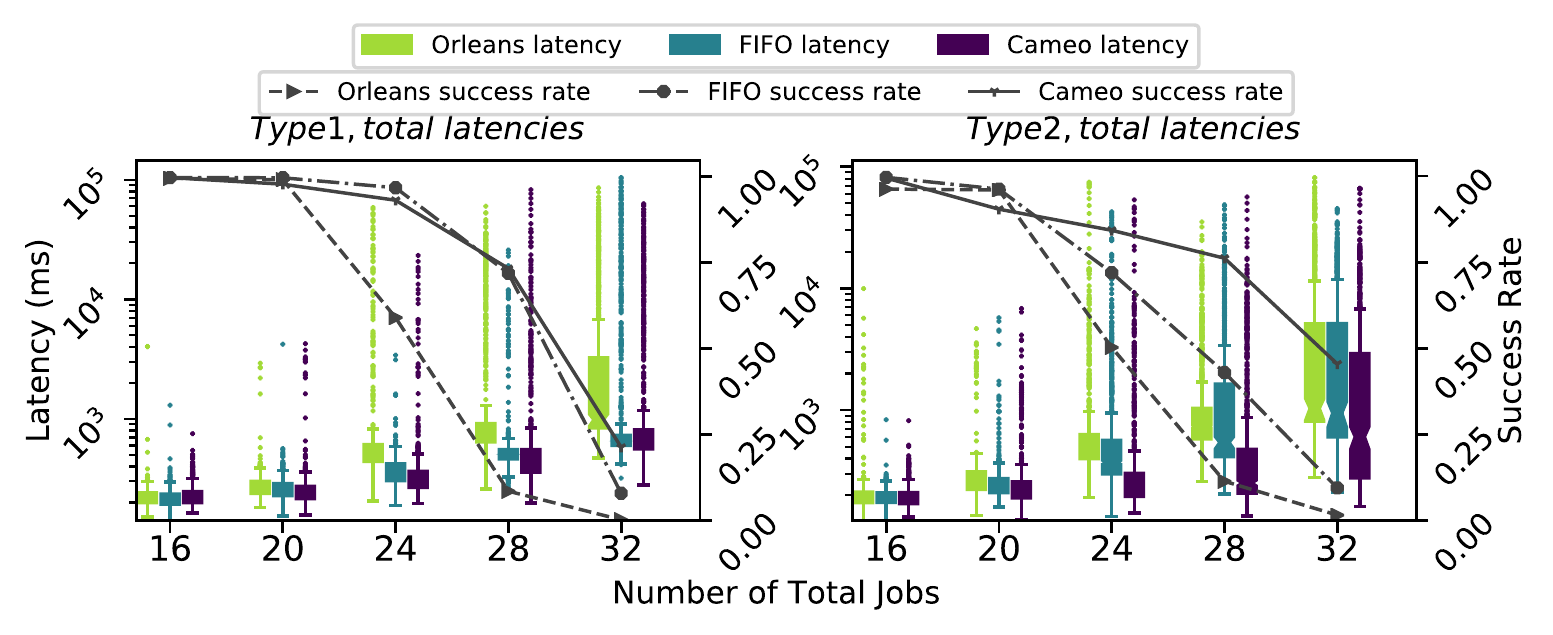}
\vspace{-0.25in}
\caption{\it Spatial Workload Variation.}
\label{exp:motivation_dataset_boxes}
\end{figure}

\noindent\textbf{Ingestion pattern from production trace.} %
Production workloads exhibit high degree of skew across data sources. In Figure~\ref{exp:motivation_dataset_boxes} we show latency distribution of dataflows consuming two workload distributions derived from Figure~\ref{fig:volume_heatmap}: Type 1 and 2. Type 1 produces twice as many events as Type 2. However, Type 2 is heavily skewed and its ingestion rate varies by 200$\times$ across sources. %
This heavily impacts operators that are collocated. 
The success rate (i.e., the fraction of outputs that meet their deadline) is only %
0.2\% and 1.5\% for Orleans and 7.9\% and 9.5\% for FIFO. \sys prioritizes critical messages,  
maintaining success rates of 21.3\% and 45.5\% respectively.%

\begin{figure}[!bt]
\centering

\includegraphics[width=.5\textwidth]{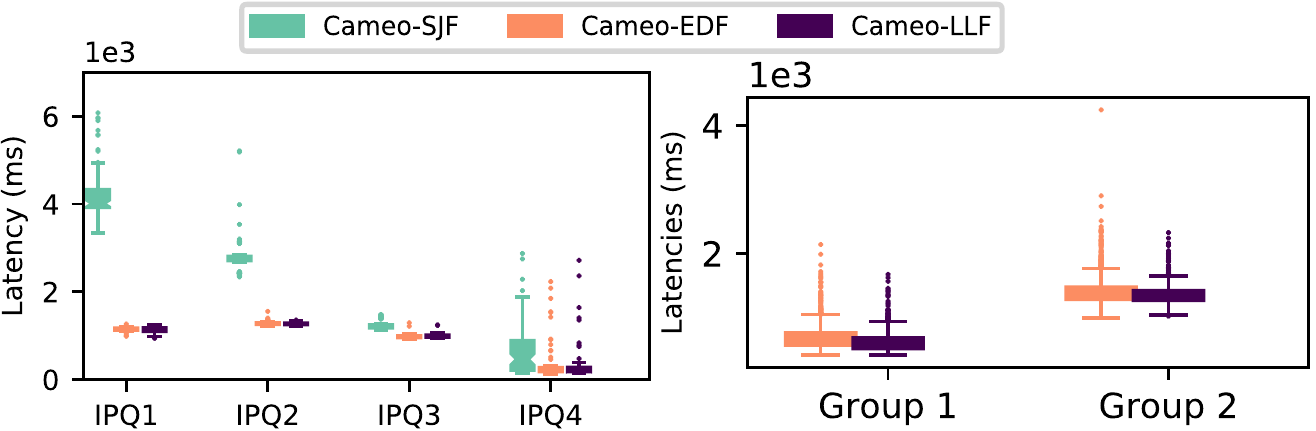}
\vspace{-0.1in}

\caption{\it %
\sys Policies. Left: Single query latency distribution. Right: Multi-Query Latency Distribution.  %
}
\vspace{-.1in}
\label{exp:real_time_strategy_comparison}
\end{figure}

\subsection{\sys: Internal Evaluation}
\label{subsec:internal_variation}

We next evaluate \sys's internal {properties}. %

{\mypar{LLF vs. EDF vs. SJF}}
 We implement three scheduling policies using the \sys context API and evaluate using   Section~\ref{subsubsec:single_job}'s workload.  
The three scheduling policies are: Least Laxity First (LLF, our default), Earliest Deadline First (EDF), and Shortest Job First (SJF). %
Figure~\ref{exp:real_time_strategy_comparison} shows  that SJF is consistently worse than LLF and EDF (with the exception of query IPQ4-- due to the lack of  queuing effect under lighter workload).
Second, EDF and LLF perform comparably.

In fact we observed that EDF and LLF produced similar schedules for most of our queries. 
This is because: i) our operator execution time is consistent within a stage, and ii)  operator execution time is $\ll$  window size. Thus, excluding operator cost (EDF) does not change schedule by  much. %

{\mypar{Scheduling Overhead}}
To evaluate \sys with many small messages, we use one thread to run a no-op workload (300-350 tenants,  1 msg/s/tenant, same latency needs). Tenants are increased to saturate throughput.

\begin{figure}[!bt]
\centering
\includegraphics[height=1.46in]{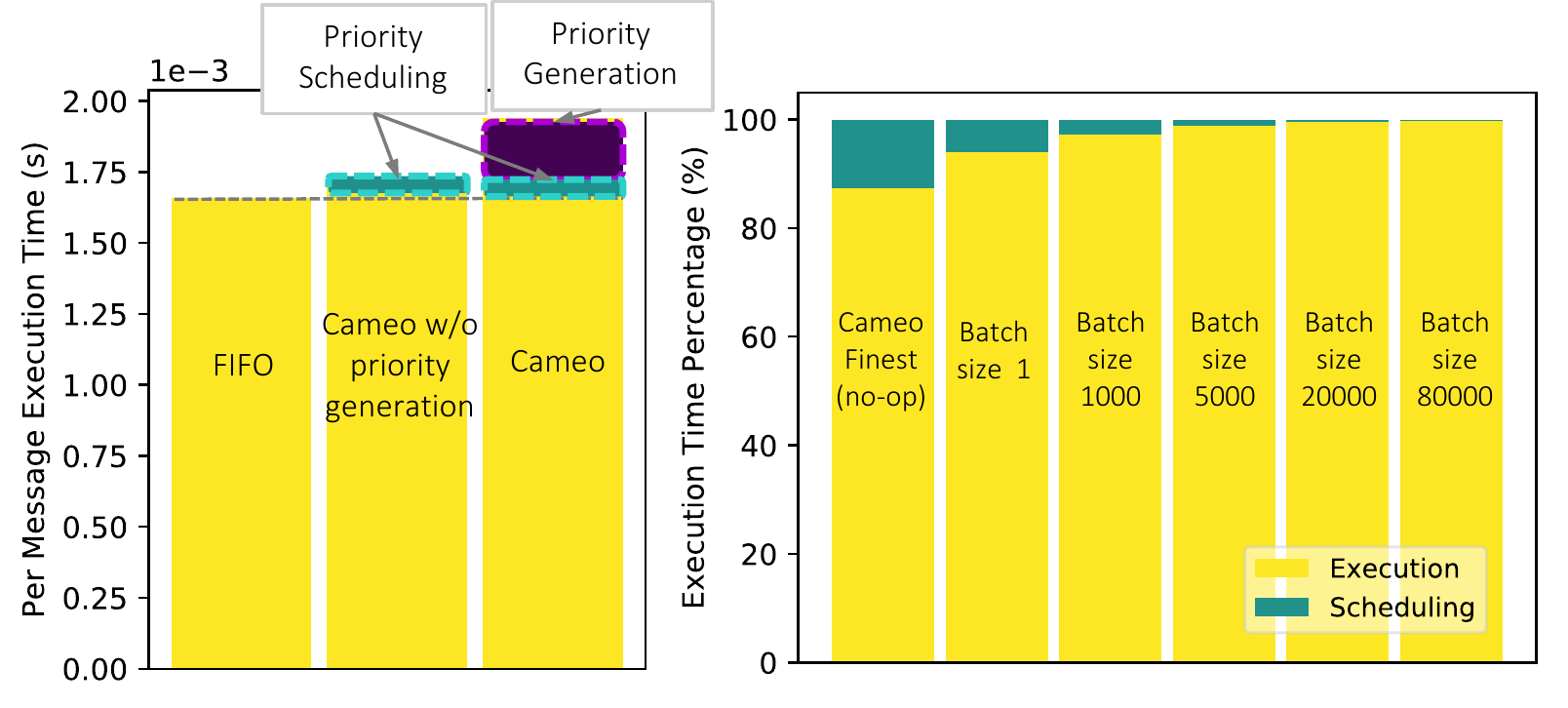}
\vspace{-0.1in}
\caption{\it %
\sys Scheduling Overhead. %
}
\vspace{-.1in}
\label{exp:scheduling_overhead}
\end{figure}

Figure~\ref{exp:scheduling_overhead} (left) shows breakdown of execution time %
(inverse of throughput) %
for three scheduling schemes: FIFO, \sys without priority generation (overhead only from priority scheduling), and \sys with priority generation and the LLF policy from Section~\ref{sec:scheduling} (overhead from both priority scheduling and priority generation). 
\sys's scheduling overhead is $<15\%$ of processing time in the worst case, comprising of $4\%$ overhead from priority-based scheduling and $11\%$ from priority generation.

In practice, \sys encloses a columnar batch of data in each message like Trill~\cite{chandramouli2014trill}. \sys's  overhead is small compared to message execution costs. In Figure~\ref{exp:scheduling_overhead} (right), under  Section~\ref{subsec:eval_setting}'s workload,  scheduling overhead is only 6.4\% of  execution time for a local aggregation operator  with batch size 1. Overhead falls with batch size. 
When \sys is used as a generalized scheduler and  message execution costs are small (e.g., with $<$ 1 ms),  we recommend tuning  scheduling quantum and message size to reduce scheduling overhead.

\begin{figure}[!bt]
\centering
\includegraphics[width=0.45\textwidth]{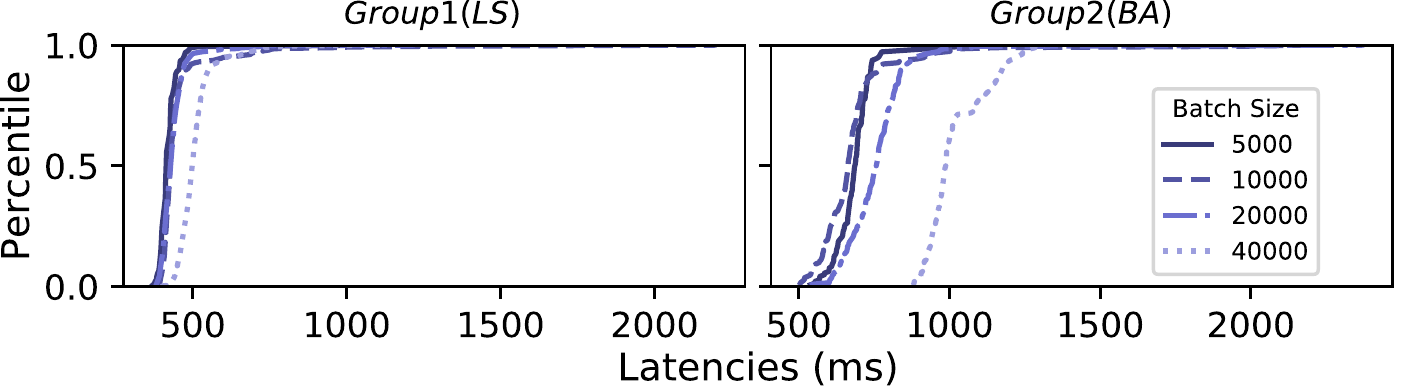}
\vspace{-0.1in}
\caption{\it Effect of Batch Size. %
}
\vspace{-0.1in}
\label{exp:scheduling_batchsize_cdfs}
\end{figure}

In Figure~\ref{exp:scheduling_batchsize_cdfs}, we batch more tuples into a message, while maintaining same overall tuple ingestion rate. In spite of decreased flexibility available to the scheduler, group 1 jobs' latency is unaffected up to 20K batch size. It degrades at higher batch size (40K), due to more lower priority tuples blocking higher priority tuples. Larger messages hide scheduling overhead, but could starve some high priority messages.%

 \begin{figure}[!bt]
\centering
\includegraphics[width=0.45\textwidth]{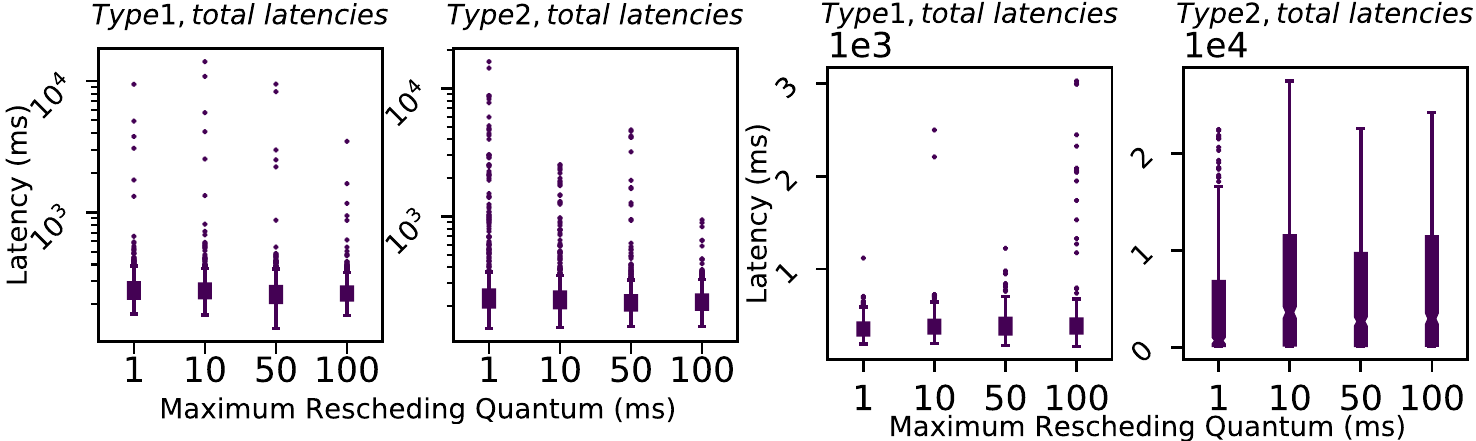}
\vspace{-0.1in}
\caption{\it Effect of Varying Scheduling Quantum. Left: Jobs Triggered By Clustered Stream Progress. Right: Jobs Triggered By Interleaved Stream Progress.}
\vspace{-0.1in}
\label{exp:rescheduling_quantum}
\end{figure}

To evaluate the effect of increasing the scheduling quantum, we evaluate \sys's performance under varying scheduling quantum (Section~\ref{subsec:architecture}) using workload described in Figure~\ref{exp:motivation_dataset_boxes}. Figure~\ref{exp:rescheduling_quantum} (left) shows the latency distribution with \textit{all} Type 1 and Type 2 jobs trigger dataflow output on the \textit{same} stream progress (e.g., 10, 20, 30s ...), whereas Figure~\ref{exp:rescheduling_quantum} (right) shows the result with jobs output triggered on \textit{interleaved} stream progress (e.g., job 1 on 10, 20, 30s etc., job 2 on 12, 22, 32s, etc.). The left figure reveals the potential benefit of using a coarser scheduling quantum, as resources are contended by many high priority messages, using finest scheduling granularity causes longer latency tail due to frequent context switches. However, a very large scheduling quantum (100ms) can hurt \sys's performance by prohibiting the scheduler from preempting low-priority operators that arrive early, creating head-of-line blocking for high priority messages.

\begin{figure}[!bt]
\centering
\begin{tabular}{cc}
\includegraphics[width=0.45\textwidth]{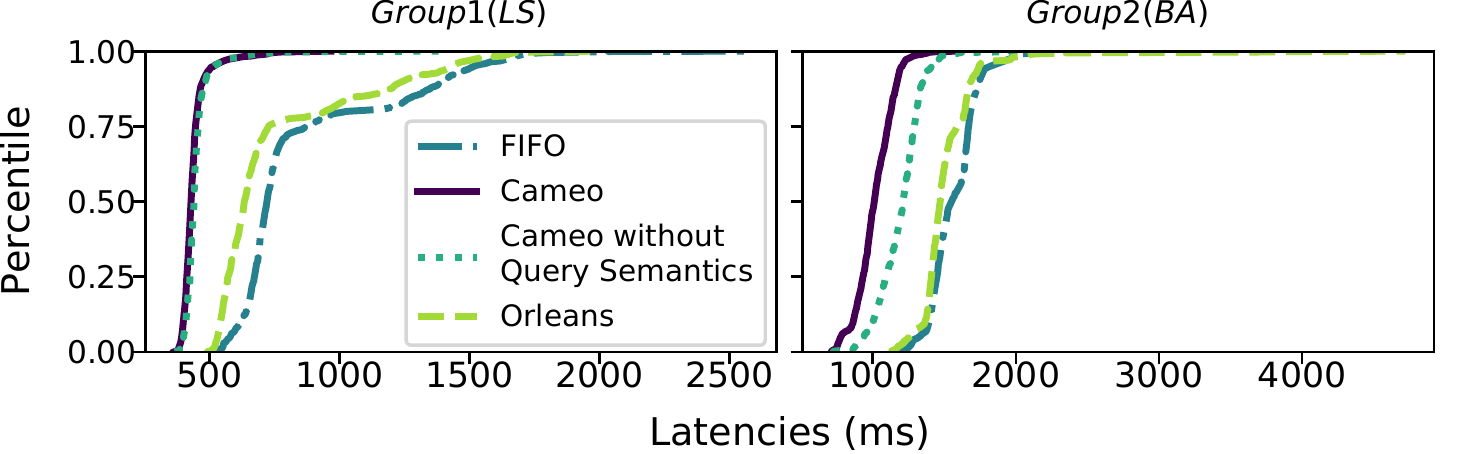}
\end{tabular}
\vspace{-0.1in}
\caption{\it Benefit of Query Semantics-awareness in \sys.
}
\vspace{-0.1in}
\label{exp:scheduling_slo_only}
\end{figure}

{\mypar{Varying Scope of Scheduler Knowledge}}
If \sys is unaware of query semantics (but aware of DAG and latency constraints), 
\sys conservatively estimates $t_{M_{F}}$ without deadline extension for window operators, causing a tighter $ddl_{M}$.  
Figure~\ref{exp:scheduling_slo_only} shows that %
\sys performs slightly worse without query semantics (19\% increase in group 2 median latency). Against baselines, \sys still reduces group 1 and group 2's median latency by up to 38\% and 22\% respectively. %
Hence, even without query semantic knowledge, \sys still outperforms Orleans and FIFO.

\begin{figure}[!bt]
\centering
\begin{tabular}{cc}
\includegraphics[width=0.45\textwidth]{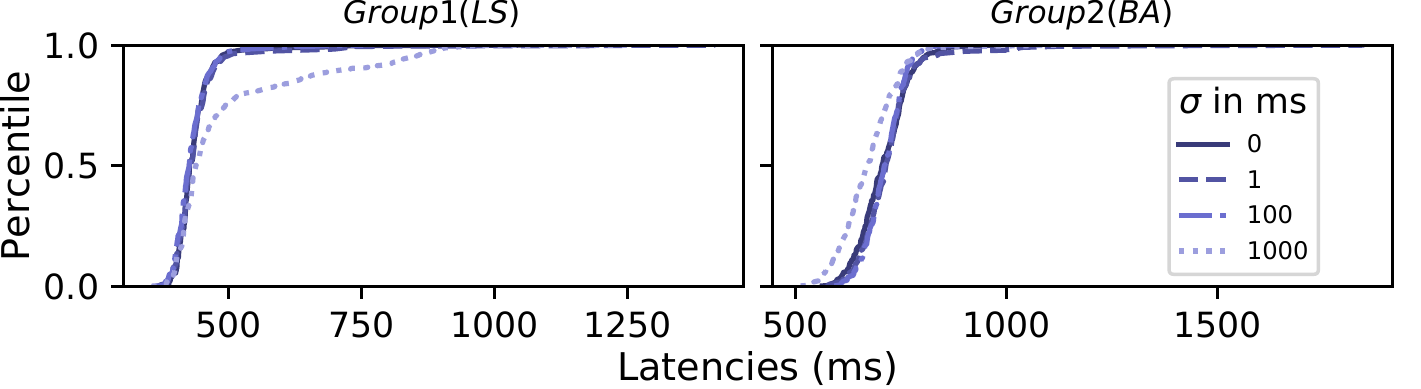}
\end{tabular}
\vspace{-0.1in}
\caption{\it Profiling Inaccuracy. Standard deviation in ms.
}
\vspace{-0.1in}
\label{exp:measurement_errors}
\end{figure}

{\mypar{Effect of Measurement Inaccuracies}}
To evaluate how \sys reacts to inaccurate monitoring profiles, we perturb measured profile costs  {($C_{O_{M}}$ from  Equation~\ref{eq:ddl-3})} by a normal distribution ($\mu$=0), varying standard deviation ($\sigma$) from 0 to 1 s. 
Figure~\ref{exp:measurement_errors} shows that when $\sigma$ of perturbation is close to  window size (1 s), latency is: i) stable at the median, and ii) modestly increases at tail, e.g., only by {55.5\%} at the {90th}  percentile. %
Overall, \sys's performance is robust when standard deviation is $\leq$ 100ms, i.e.,  when measurement error is reasonably smaller than output granularity.

\section{Related Work}
\mypar{Streaming system schedulers}
The first generation of Data Stream Management Systems (DSMS)~\cite{abadi2003aurora, cherniack2003scalable}, 
such as Aurora~\cite{carney2003operator}, %
Medusa~\cite{balazinska2004load} and Borealis~\cite{abadi2005design}, use QoS based control mechanisms with load shedding to improve query performance at run time. %
These are either centralized (single-threaded)~\cite{carney2003operator}, or  distributed~\cite{balazinska2004load,abadi2005design} but do not  handle timestamp-based priorities for partitioned operators. %
TelegraphCQ~\cite{chandrasekaran2003telegraphcq} orders input tuples before query processing
~\cite{avnur2000eddies, raman1999online}, while \sys addresses operator scheduling within and across query boundaries. Stanford's  STREAM~\cite{motwani2003query} uses chain scheduling~\cite{babcock2003chain} to minimize memory footprints and optimize query queues, but  assumes all queries and  scheduler are execute  in a single-thread. More recent works in streaming engines propose  operator scheduling algorithms for query throughput~\cite{amini2006adaptive} and latency~\cite{li2009earliest, gu2012deadline}. %
Reactive and operator-based policies include  \cite{amini2006adaptive, li2009earliest}, while \cite{gu2012deadline} assumes arrivals are periodic or Poisson---however, these works do not build a framework (like \sys), nor do they handle per-event semantic awareness for stream progress. %

Modern stream processing engines such as Spark Streaming~\cite{sparkstreaming},  Flink~\cite{flink}, Heron~\cite{heron}, MillWheel~\cite{akidau2013millwheel}, Naiad~\cite{murray2013naiad}, Muppet~\cite{lam2012muppet}, Yahoo S4~\cite{neumeyer2010s4}) %
do not include \textit{native} support for multi-tenant SLA optimization. These systems also rely on coarse-grained resource sharing~\cite{storm_multitenant} or third-party resource management systems such as YARN~\cite{yarn} and Mesos~\cite{apache_mesos}.%

\mypar{Streaming query reconfiguration} Online reconfiguration has been studied extensively~\cite{hirzel2014catalog}. Apart from Figure~\ref{fig:related_work_venn}, prior work addresses  operator placement~\cite{pietzuch2006network, garefalakis2018medea}, load balancing~\cite{kleiminger2011balancing, loesing2012stormy}, state  management~\cite{castro2013integrating}, policies for scale-in and scale-out \cite{heinze2014latency, heinze2014auto, heinze2015online, lohrmann2015elastic}. Among these are techniques to  address latency requirements of dataflow jobs~\cite{heinze2014latency, lohrmann2015elastic}, and ways to improve vertical and horizontal elasticity of dataflow jobs  in  containers~\cite{wu2015chronostream}.  The performance model in \cite{li2015supporting} focuses on dataflow jobs with latency constraints, 
while we focus on interactions among operators. Online elasticity was targeted by System S~\cite{schneider2009elastic, gedik2014elastic}, StreamCloud~\cite{gulisano2012streamcloud} and TimeStream~\cite{qian2013timestream}. 
Others include~\cite{floratou2017dhalion,kalim2018henge}. %
Neptune \cite{garefalakis2019neptune} is a proactive scheduler to suspend low-priority batch tasks in the presence of streaming tasks. Yet, there is no operator prioritization  \textit{within} each application. Edgewise~\cite{fu2019edgewise} is a  queue-based  scheduler based on operator load but not query semantics. 
All these works are orthogonal to, and can be treated as pluggables in,  \sys. 

\mypar{Event-driven architecture for real-time data processing} This area has been popularized by 
the resource efficiency of serverless  architectures~\cite{azure_function,aws_lambda,google_cloud_functions}.
Yet, recent proposals~\cite{moritz2018ray,statefun, kumar2020amber,akhter2019stateful} for stream processing  atop event-based frameworks do not support performance targets for streaming queries. %

\section{Conclusion}

We proposed \sys, a fine-grained scheduling framework for distributed stream processing. %
To realize flexible per-message scheduling, we implemented a stateless scheduler, contexts that carry important static and dynamic information, and mechanisms to derive laxity-based priority from contexts. Our experiments with real workloads, and on Microsoft Azure, showed that \sys achieves %
$2.7\times-4.6\times$  
lower latency than competing strategies 
and incurs overhead less than $6.4$\%. %

\section*{Acknowledgements}
We thank Matei Zaharia and our anonymous referees at NSDI 2020 for their reviews and help with improving the paper. We thank Kai Zeng for providing feedbacks for initial ideas. This work was supported in part by the following grants: NSF IIS 1909577, NSF CNS 1908888, NSF CNS 1319527, NSF CNS 1838733, a Facebook faculty research award, the Helios project at Microsoft~\cite{potharaju13helios}, and another generous gift from Microsoft. Shivaram Venktaraman is also supported by the Office of the Vice Chancellor for Research and Graduate Education with funding from the Wisconsin Alumni Research Foundation. We are grateful to the Cosmos, Azure Data Lake, and PlayFab teams at Microsoft.

\normalem

\bibliographystyle{plain}
\bibliography{paper}
\end{document}